\documentclass[aps,superscriptaddress,twocolumn,nofootinbib]{revtex4}

\usepackage{natbib}
\usepackage{amsfonts}
\usepackage{pslatex}
\usepackage{ae,aecompl}
\usepackage{epsfig}
\usepackage{graphics}
\usepackage{graphicx}
\usepackage{amsmath}
\usepackage{amssymb}
\usepackage{pifont}
\usepackage{psfrag}
\usepackage{subfigure}
\usepackage{float}
\usepackage{color}
\usepackage{booktabs}  
\usepackage{multirow}
\usepackage{comment}
\usepackage{dcolumn}

\graphicspath{{./FIGURE/}}

\newcommand{\dprima}[2]{\frac{ \partial{#1}}{ \partial{#2}}}

\newcommand{\dparsec}[3]{\frac{ \partial^2{#1}}{ \partial{#2} \partial{#3}}}
\newcommand{\diff}{\mathrm{d}}
\newcommand{\E}[1]{\mathbb{E}\left[ {#1}\right]} 
\newcommand{\hist}{\cal{H}} 
\newcommand{\like}{{\cal{L}}} 
\newcommand{\ndist}{\cal{N}} 

\DeclareMathOperator{\SMA}{SMA}

\newcommand{\N}{\mathbb{N}}

\newcommand{\abs}[1]{\lvert #1 \rvert} 

\newcommand{\kerde}{\phi_{\mbox{\tiny{\textbf{DE}}}}}

\newcommand{\de}{\mbox{\tiny{\textbf{DE}}}}
\newcommand{\pla}{\mbox{\tiny{\textbf{PL}}}}

\newcommand{\AIC}{\mbox{AIC}}
\newcommand{\BIC}{\mbox{BIC}}
\newcommand{\RL}{\mbox{RL}}
\newcommand{\NC}{\mbox{\tiny{NC}}}
\newcommand{\C}{\mbox{\tiny{C}}}

\begin{document}

\title{Modeling FX market activity around macroeconomic news: \\a Hawkes process approach}
\author{Marcello Rambaldi}
\affiliation{Scuola Normale Superiore, Piazza dei Cavalieri 7, Pisa, 56126, Italy}
\author{Paris Pennesi}
\affiliation{eFX Quantitative Trading, HSBC Bank, 8 Canada Square, London E14 5HQ, UK}
\author{Fabrizio Lillo}
\affiliation{Scuola Normale Superiore, Piazza dei Cavalieri 7, Pisa, 56126, Italy}
\affiliation{Dipartimento di Fisica e Chimica, Universit\`a degli Studi di Palermo, Viale delle Scienze Ed. 18, Palermo, 90128, Italy}
\affiliation{Santa Fe Institute, 1399 Hyde Park Road, Santa Fe, NM 87501, USA}

\begin{abstract}
We present a Hawkes model approach to foreign exchange market in which the high frequency price dynamics is affected by a self exciting mechanism and an exogenous component, generated by the pre-announced arrival of macroeconomic news. By focusing on time windows around the news announcement, we find that the model is able to capture the increase of trading activity after the news, both when the news has a sizeable effect on volatility and when this effect is negligible, either because the news in not important or because the announcement is in line with the forecast by analysts. We extend the model by considering non-causal effects, due to the fact that the existence of the news (but not its content) is known by the market before the announcement. 
\end{abstract}

\maketitle

\section{Introduction}
\label{intro}

Complex system dynamics is characterized by a subtle interplay between endogenous and exogenous effects. Financial markets are, in this respect, paradigmatic. The dynamics of asset prices, in fact, is affected by the arrival of exogenous news, which modifies the valuation of the fair price of the asset, and by the arrival of endogenously generated events (e.g. trades, orders, and price changes) in reaction to past events \cite{Engle1993,Birz2011,Gross2011}. Disentangling and quantifying the relative importance of these two drivers have been debated widely in the last half century, since the formalization of the Efficient Market Hypothesis. There is a growing empirical evidence that a significant fraction of market activity\footnote{In this paper we will use the term {\it market activity} to describe the rate of price changes per unit time, irrespectively on the size, similarly to what investigated in \cite{filimonov2012quantifying,bouchaud_hawkes}} and volatility is explained by the endogenously generated trading activity.  For example, large price movements (jumps) are only partly explained by public news \cite{Cutler1989,joulin2008stock}.

The debate about the endogenous and exogenous component of price dynamics has recently received a new impulse thanks to the application of Hawkes processes to the modeling of financial data \cite{hewlett2006,Bowsher2007,bauwens2009_review,embrechts2011,toke2011,bacry2011,bacry2012non,bormetti_cojumps}. Hawkes processes \cite{hawkes1971}, originally introduced in seismology, describe a point (counting) process where the intensity is not constant but depends on the past history of the counting process, weighted by a suitably chosen kernel. Due to the natural interpretation of a Hawkes process as a branching process, in terms of immigrant baseline events and offspring events, it is direct to interpret the integral of the kernel as the fraction of the activity due to the endogenous self excitation. The empirical analysis on equity and future data has indicated \cite{filimonov2012quantifying,bouchaud_hawkes,filimonov2013,hardiman2014} that, according to Hawkes modeling, a very large fraction of the market activity is explained by the endogenous component. Markets seem to be very close to a "critical" state where all the dynamics is endogenously generated (see \cite{filimonov2012quantifying,bouchaud_hawkes} for more details). 

However in some situations the arrival of exogenous news has typically a very large impact on the trading activity. An example, which is the one investigated in the present paper, is the impact of macroeconomic news on foreign exchange (FX) markets. This large impact is due to the fact that macroeconomic news (unemployment rate, GDP data, central banks intervention, etc.) affects significantly the valuation of the economy of a country and therefore of its currency. Moreover these announcements are scheduled in advance and are monitored with great attention by FX market participants. 

The impact of exogenous news on asset prices has been investigated by a number of authors starting with the pioneering work of \cite{Cutler1989}. In Ref. \cite{mitchell1994impact} authors analysed the connection between the daily number of news announcements reported by Dow Jones \& Company and aggregate measures of market activity such as trading volume and market returns. Avellaneda and Lipkin \cite{avellaneda2003market} studied the phenomenon of stock pinning observed at the expiration date of the corresponding options. Refs. \cite{joulin2008stock} and \cite{gross2011machines} explored the role of news in high-frequency volatility and order book dynamics. A more extensive survey of the literature on news impact can be found in \cite{lillo2014news}.

The effects of scheduled macroeconomic announcements on the FX market have been the object of several works. Ederington and Lee \cite{ederington1993markets} found that this sort of announcements is responsible for a large part of the volatility pattern observed in these markets. Ref. \cite{andersen1998deutsche} provided a characterization of the volatility in the DM/USD rate, separating contributions of intraday activity patterns, macroeconomic announcements, and volatility persistence.
More recently, Bauwens et al. \citep{Bauwens20051108} have a investigated volatility dynamics around news announcements on high-frequency EUR/USD data in the framework of ARCH-type models. These works focus either on large time horizons or choose a discrete time setting.

In this paper we focus on modeling the dynamic evolution in physical (continuous) time of market activity before and after a macroeconomic news announcement. We consider the rate of change in the best quotes as a proxy of market activity, and we take a point process perspective. Our main original contribution is a Hawkes model of the FX market activity where one of the kernels describes the (causal and possibly non-causal) effect of the macro news on the market activity. The model has thus two Hawkes kernels, one for the endogenous and one for the exogenous component. We find that the description of the market activity improves significantly when one introduces the exogenous kernel. It is important to stress that most of the work done so far concerning Hawkes processes in financial data has considered stationary processes. In fact, when non stationarities are present (e.g. the intraday pattern), one typically tries to remove them, for example by deseasonalizing the data. However in some cases, non stationarities are important and bring relevant economic information that one wants to model. The example considered in this paper is the market activity around macro announcements. Our approach demonstrates how it is possible to reconcile non stationarities like news announcements in the Hawkes-processes framework, that has proven to be a versatile tool for modeling of high-frequency market dynamics in physical time.

The paper is organized as follows. In Section \ref{data} we present our dataset of prices and news and the main variables introduced to characterize the news. In Section \ref{sec:Hawkes} we show how Hawkes models are able to describe the dynamics of FX rates when only the endogenous component is considered. Section  \ref{sec:Hawkes_news} presents our modeling approach, where a new component of the Hawkes process is added, taking into account the effect of macro news. In Section \ref{sec:noncausal} we present an extension of the model where a non-causal kernel is added, taking into account the fact that the existence of the announcement (but not its content) is known in advance. Finally, in Section \ref{sec:conclusions} we present some conclusions. In Appendix \ref{sec:like_appendix} we give the details of maximum log-likelihood estimation. Appendix \ref{sec:comparison_appendix} provides further material on the comparison between the model with the new exogenous component and the one without it.

\section{Data}\label{data}

We investigated EBS live data from January 1, 2012 to December 18, 2012, a total of 353 days, for three pairs of currencies, namely EUR/USD, EUR/JPY and USD/JPY. The time resolution of the database is 100ms and we used the \textit{ebsMarketReferenceTime} as the reference time. Since the time discretization might affect the estimation of the Hawkes process, as customary done, we randomize the times by subtracting from each time stamp expressed in second a random number uniformly distributed in (0,0.1] s.

We filter the data by considering the events where either the best ask or the the best bid change. Although FX trading is active 24h a day 7 days a week, the activity on Saturdays and Sundays recorded in our dataset is almost negligible and therefore we excluded them.  This left us with 250 days. FX trading activity displays an high intraday seasonality and therefore we decided to restrict our work to London core trading hours (07:30-16:30), where the highest activity is observed. Table \ref{tab:summary_fx} shows the number of bid or ask changes observed during the London core trading hours for the three rates. Notice that we have divided the sample in two, depending on the tick size\footnote{The tick size is the discretization step of price.}. In fact EBS increased the tick size of a factor five on September 23, 2012 and, as it can be seen from the Table, this fact has reduced the number of best quote changes by a factor two or three, depending on the rate.

\begin{table}
\small
\begin{tabular}{lcrc}
  \toprule[1pt]
&\# of days & Events per   & Avg. quote  \\
&                 & day (Avg.)   & duration (s) \\
\midrule
  &\multicolumn{3}{c}{Small tick size}\\
 \midrule
EUR/USD &190 & 58,902 & 0.55\\
EUR/JPY &190 & 34,857 & 0.91\\
USD/JPY &190& 24,477 &1.32 \\
\midrule
&\multicolumn{3}{c}{Large tick size}\\
\midrule
EUR/USD &60 & 20,229 & 1.60\\
EUR/JPY &60 &  19,012 &1.70\\
USD/JPY &60& 7244 & 4.47\\
\bottomrule[1pt]
\end{tabular}
\caption{Summary statistics for the FX dataset during London core trading hours (07:30-16:30), before and after the tick size increase. Events refer to cases when either the bid or the ask change (with a time resolution of 100ms).}
\label{tab:summary_fx}
\end{table}

The FX market is strongly affected by macro announcements. In this paper we are interested in modeling the price dynamics around these announcements. In particular here we focus on announcements whose publishing is scheduled well in advance, such as  Central Banks rate decisions or announcements of unemployment or GDP data. We obtained the list of all economic announcements of 2012 from the website \emph{www.dailyfx.com}. The news data report date and time of the news release, the currency which is most affected by the news, a brief description of the news, an estimate of the importance (high, medium, low), and, where available, the expected (Forecast) and actual value. The Forecast value is an average of analysts' expectations.  Not all the news in the dataset concern the publication of an indicator which can be quantified by a number. An example are press conferences of central banks' governors. Since there is no numeric value associated to them, there is not any forecast either. Since we treat only the EUR/USD, EUR/JPY and USD/JPY rates, we retained only the news that are linked to one of these currencies. In addition, in this work we consider only news that are rated Medium or High importance, so we discarded those rated Low. Quite often, related indicators are announced at the same time. Since we are concerned with the impact of a news before and after its announcement, we kept only the most relevant news for each distinct time. The total of the news retained after this selection is 723. The totals divided per currency and importance are reported in Table \ref{tab:tab_news_number}. The much smaller number of JPY news is due to the fact that we focus on London core trading hours, and most JPY news are released outside this time interval.

\begin{table}[ht]
\begin{center}
\begin{tabular}{lrrr}
\toprule
& HIGH & MEDIUM &TOTAL\\
\midrule
EUR 	& 130  	& 293 & 423\\
USD 	& 107 	& 186 & 293\\
JPY 	&      1  	&      6 &     7\\
\midrule
TOTAL	& 238	& 485 & 723\\
\bottomrule 
\end{tabular}
\caption{Number of news affecting each currency disaggregated by importance. }
\label{tab:tab_news_number}
\end{center}
\end{table}

\subsection{Activity jumps and news surprise}\label{sec:news_surprise}

In order to characterize the news, we introduce two additional metrics. The first one measures how much the news has impacted the trading activity, while the second one measures the degree of surprise of the news, i.e. the difference between the forecast and the actual value. 

To be more specific, for each one minute interval $i$ we compute the number $N_i$ of changes in any of the best quotes, and we compare it with the Simple Moving Average ($\SMA_n$) of the same quantity over the previous $n$ one minute intervals. Since the events in minute $i$ do not enter in the calculation of $\SMA_n(i)$, we consider it as a measure of the expected activity for the interval $i$. Similarly to what is done in price jumps detection \cite{joulin2008stock,bormetti_cojumps}, we define the {\it impact} $\theta_i$ as the ratio between the observed number of events at $i$ and the expected number of events according to SMA, namely 
\begin{equation}
\theta_i=\frac{N_i}{\SMA_n(i)}
\end{equation}
We shall use the quantity $\theta_i$ for two purposes. First we use it to identify "jumps" in the market activity and to answer the question of how frequently a high activity is associated with a recent news. Second we shall use is to answer the reverse question, i.e. how large is $\theta_i$ after a macro news and to classify news according to their effect on the trading activity.

The second metric measures the difference between expectations and actual values announced by the news. For this reason we call it "surprise" of the news. To compute it, we can only consider news for which the Forecast value is given. In the end, starting with $723$ news, we are left with a total of $559$ news or about $77\%$ of the original sample.

For each news we have the Actual and the Forecast value, which we denote $I_A$ and $I_F$ respectively. The measures of surprise that we use are the absolute surprise
\begin{equation}
\label{eq:abs_surprise}
S_{abs}=\abs{I_A-I_F},
\end{equation}  
and the relative surprise
\begin{equation}
\label{eq:rel_surprise}
S_{rel}=\frac{\abs{I_A-I_F}}{I_F}\times 100.
\end{equation}  
In principle one could distinguish between "good news" and "bad news", i.e. if the indicator was better or worse than expected. However this is difficult to do in an unsupervised way, since a positive increment is good or bad depending on the type of news. Therefore we chose to use the absolute value of the difference. 
The choice between the absolute and the relative surprise is not immediate. On one hand, in fact, indicators span a wide range of values, some are of the order of unity other are given in thousands or in percentage and hence the most reasonable choice seems the relative surprise. On the other hand, for some indicators, such as unemployment rate, that are given in percentage, even a small absolute change might be significant. For these reasons we use the absolute surprise for indicators given in percentage and the relative surprise in all other cases. In the following, the notation $S$ denotes this combined measure.

\begin{figure}[t]
\centering
\includegraphics[width=0.4\textwidth]{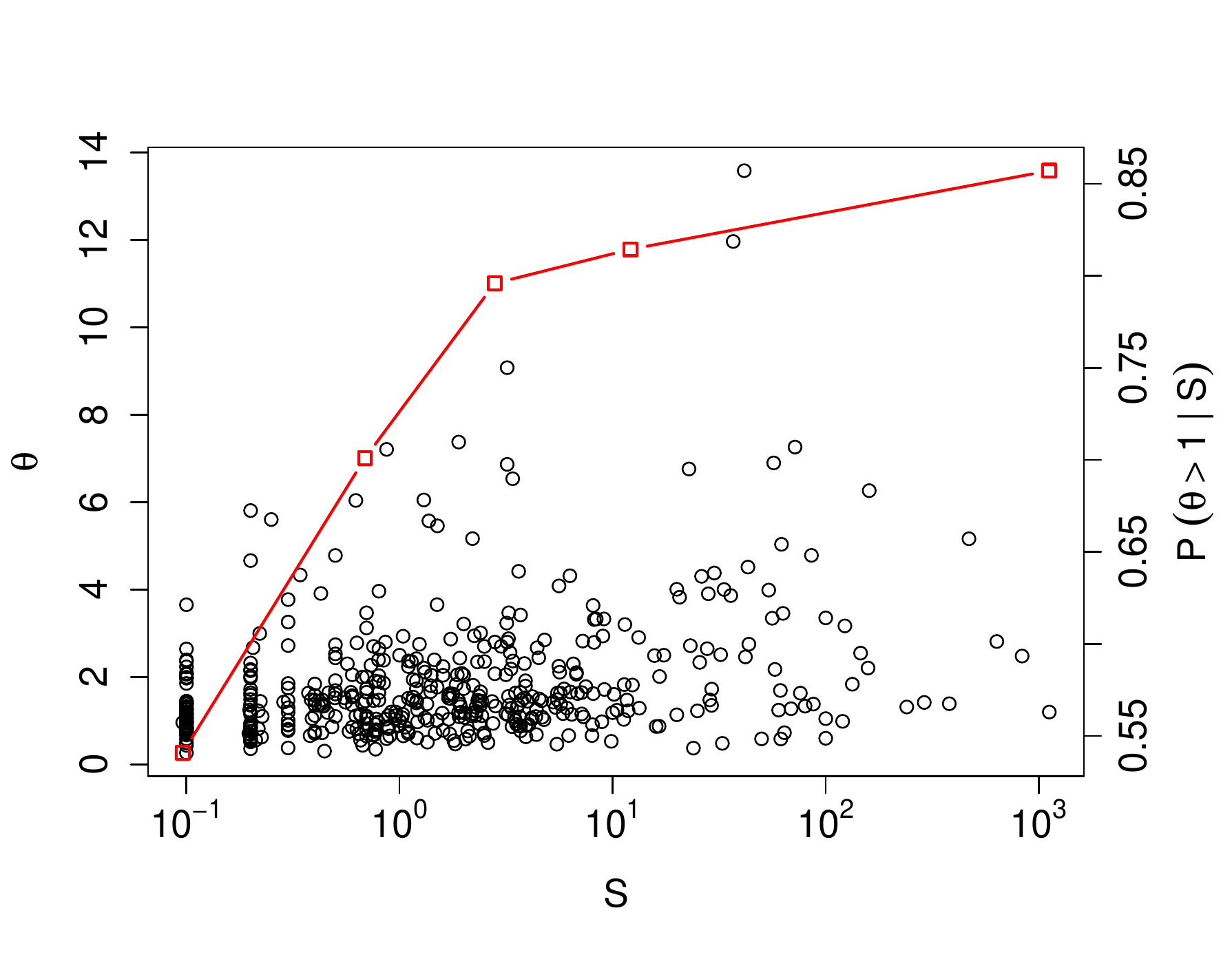}
\caption{(Color online). Scatterplot of $\theta$ against our measure of news surprise $S$ (left y axis). The figure refers to EUR/USD data for $\theta$. The red line (right y axis) is an estimation of $P(\theta>1|S)$. For reference, when $S=0$ this probability is $0.73$}
\label{fig:surprise_scatter}
\end{figure}

The correlation between surprise and $\theta$ is surprisingly low and the scatterplot is very noisy\footnote{One could try to reduce this noise by normalizing the surprise indicator by the historical standard deviation, thus taking into account the typical uncertainty of the surprise of each news. This however would require much longer time series in order to have enough observations for each type of news.}. By restricting to Medium and High importance news, the Spearman correlation coefficient between surprise and $\theta$ is $0.34$ and different from zero in a statistically significant way. This effect comes mainly from the High importance news, for which the Spearman correlation raises to $0.62$. Figure \ref{fig:surprise_scatter} shows the scatterplot of $\theta$ vs. the surprise for EUR/USD for Medium and High importance news. 
%
This relation is confirmed by $P(\theta>1|S)$, which is the probability that $\theta$ is larger than one conditional to a given value of the surprise (see  Figure \ref{fig:surprise_scatter}).  We observe that this probability is systematically larger than $1/2$, indicating that news trigger an increase in activity and this increase depends on the surprise of the news. For EUR/USD and EUR/JPY we obtain an almost monotone trend, while USD/JPY presents a noisier pattern. It seems that even this very simple measure of surprise, that does not take into account the specificity of each indicator, is nevertheless able to capture some relevant information.

\section{Hawkes models of FX market activity}\label{sec:Hawkes}

In this Section we present the baseline model for market activity. It consists of a Hawkes process with a constant baseline intensity and a self-excitation mechanism describing the endogenous effect \cite{filimonov2012quantifying,bouchaud_hawkes}. In the next Section we will introduce also an exogenous news term.
Hawkes processes are powerful models for describing the arrival of events which cluster in time. In one dimension they model a counting process $N_t$ in continuous time with an intensity (i.e. probability per unit time of observing an event) described by
\begin{equation}
\label{eq:hawkes_def}
\lambda(t)=\mu+\int_{-\infty}^t \phi(t-s) dN_s
\end{equation}
where $dN_s= \sum_{t_i<s} \delta (s-t_i)$ is a mixture of Dirac measures, $t_i$ being the the time at which $i$th event occurred. $\mu$ is the constant baseline intensity and the kernel $\phi(\tau)\ge 0$ describes the effect of past counts to the current intensity. The process is stationary if $n\equiv\int_{0}^\infty \phi(\tau)d\tau<1$. As described in \cite{bremaud2001hawkes,bouchaud_hawkes}, the process can be stationary in the critical case $n=1$ if $\mu=0$. If $n<1$, it is related to $\mu$ through the relation 
\begin{equation}
{\Lambda} \equiv \E{\lambda(t)}=\frac{\mu}{1-n}.  
\end{equation}
The quantity $n$ is the fraction of the average rate explained by the self-exciting mechanism.

The Hawkes process is characterized by the functional form of the kernel $\phi(t)$. In empirical applications, one can choose a priori a functional form for the kernel, dependent on some parameters, and then use maximum likelihood method to determine the best fitting parameters' values \footnote{Another parametric method was recently proposed in \cite{daFonseca2014hawkes}.}. A non-parametric approach was recently developed by \cite{bacry2012non}. The log-likelihood can be computed analytically (see Eq. \eqref{eq:like_pp_univ} in Appendinx A), though numerical optimization techniques are necessary for maximization. We note here that definition \eqref{eq:hawkes_def}, which is standard in the literature, assumes that the process starts infinitely back in the past. Thus, an approximation is made when the likelihood is evaluated on a finite sample. The quality of the approximation improves as the sample size increases. 

In our application, for the endogenous kernel we use either a double exponential\footnote{We have also considered the single exponential case. In agreement with the results of \cite{bouchaud_hawkes} for other financial assets, we find that FX data are very poorly described by Hawkes processes with a single exponential kernel.} or a power-law decaying kernel.
The double exponential kernel $\kerde(t)$ is defined as
\begin{equation}
\label{eq:kernel_uni_de}
\kerde(t)=\alpha_A e^{-\beta_A t}+\alpha_B e^{-\beta_B t},
\end{equation}
\noindent where $\alpha_q$ and $\beta_q$ ($q=A,B)$ are the constant parameters of the model. Here two different time scales are present and their relative weight is controlled by the amplitudes $\alpha$. For definiteness we choose $\beta_A\ge\beta_B$, i.e. the $A$ subscript refers to the shorter time scale and $B$ to the longer one. 

The power-law kernel we use is in fact not a truly power-law decaying function, but instead, for computational reasons, we use an exponential sum that carefully approximates a function with a power law tail. In particular we adopt the same specification of \cite{bouchaud_hawkes}, which reads
\begin{equation}
\label{eq:bouchaud_kernel}
\phi_{\pla}(t)= \frac{n}{Z} \lbrace \sum_{k=0}^{M-1} (a_k)^{-p} e^{-\frac{t}{a_k} }-S e^{ -\frac{t}{a_{-1}} } \rbrace
\end{equation}
where
\begin{equation}
a_k=\tau_0 m^k.
\end{equation}
The parameters are $n$, $p$ and $\tau_0$. $M$ is the number of exponential terms and $m$ is a scale parameter.  We fix $M=15$ and $m=5$ as in \cite{bouchaud_hawkes}. The parameter $m$ controls how well a single exponential approximates the power law. If one wishes the approximation to be valid over many orders of magnitude, high values of $m$ and $M$ are necessary. $Z$ and $S$ are constants such that $\phi_{\pla}(0)=0$ and $\int_0^\infty \phi_{\pla} (t) \diff t =n$. This kernel has a smooth cut-off at short lags, provided by the negative exponential, whose time scale is regulated by $\tau_0$ once $m$ is fixed. This approximated power law kernel presents an exponential cut-off at large $t$. This difference with respect to a true power law is however irrelevant, in fact, for our choice of $M$ and $m$, it manifests itself only for $t$ larger than $\approx 10^9$s, a value much larger than the duration of a trading day. This approximation of the power law in terms of an exponential sum allows the log-likelihood to be computed recursively, reducing the computational cost form $O(N^2)$ to $O(N)$, where $N$ is the sample size. This is particularly relevant given the large number of data available. We refer to Appendix \ref{sec:like_appendix} for further details on the log-likelihoods functions. 

\begin{figure}[ht]
\centering
\includegraphics[width=0.35\textwidth]{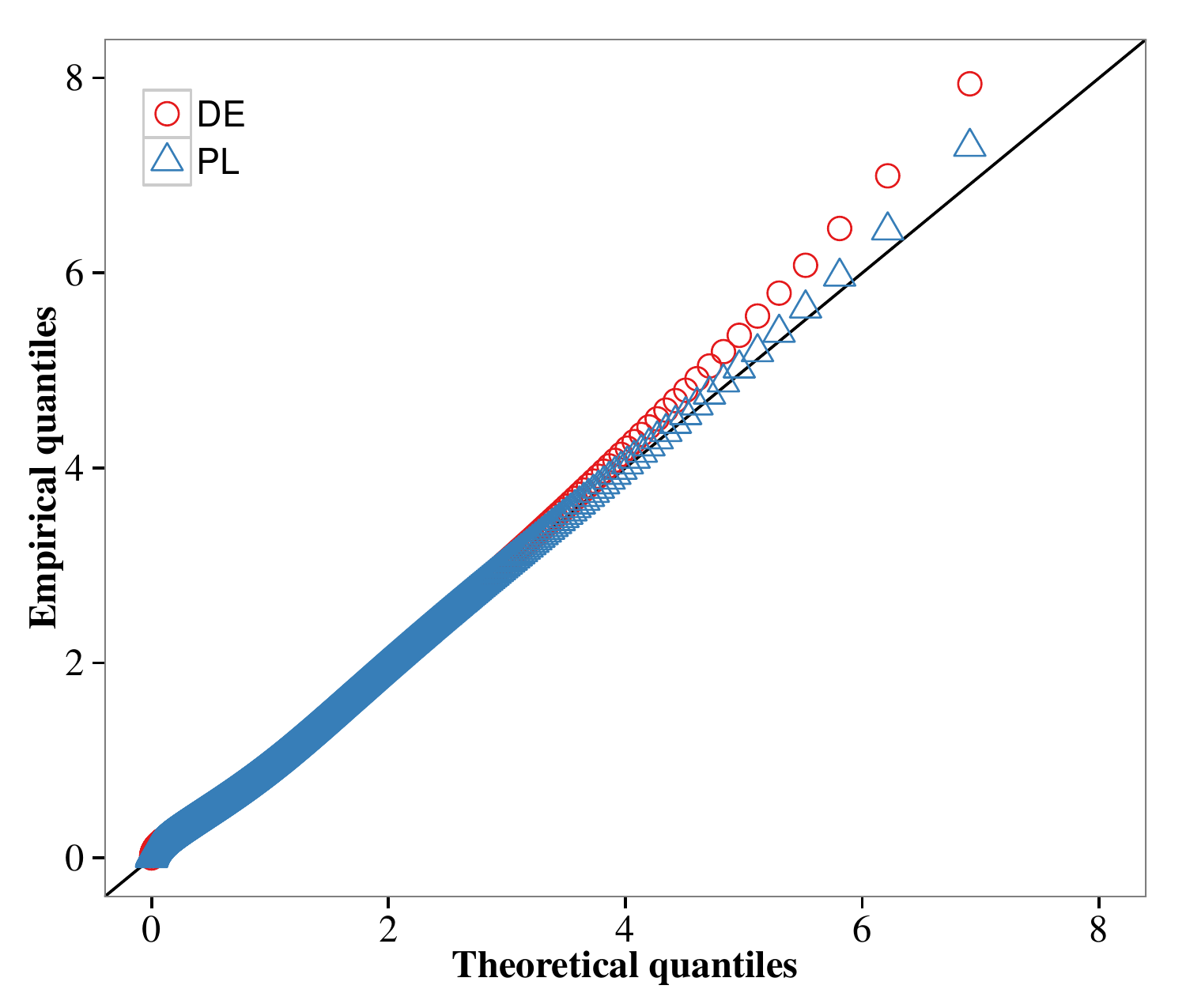}
\includegraphics[width=0.35\textwidth]{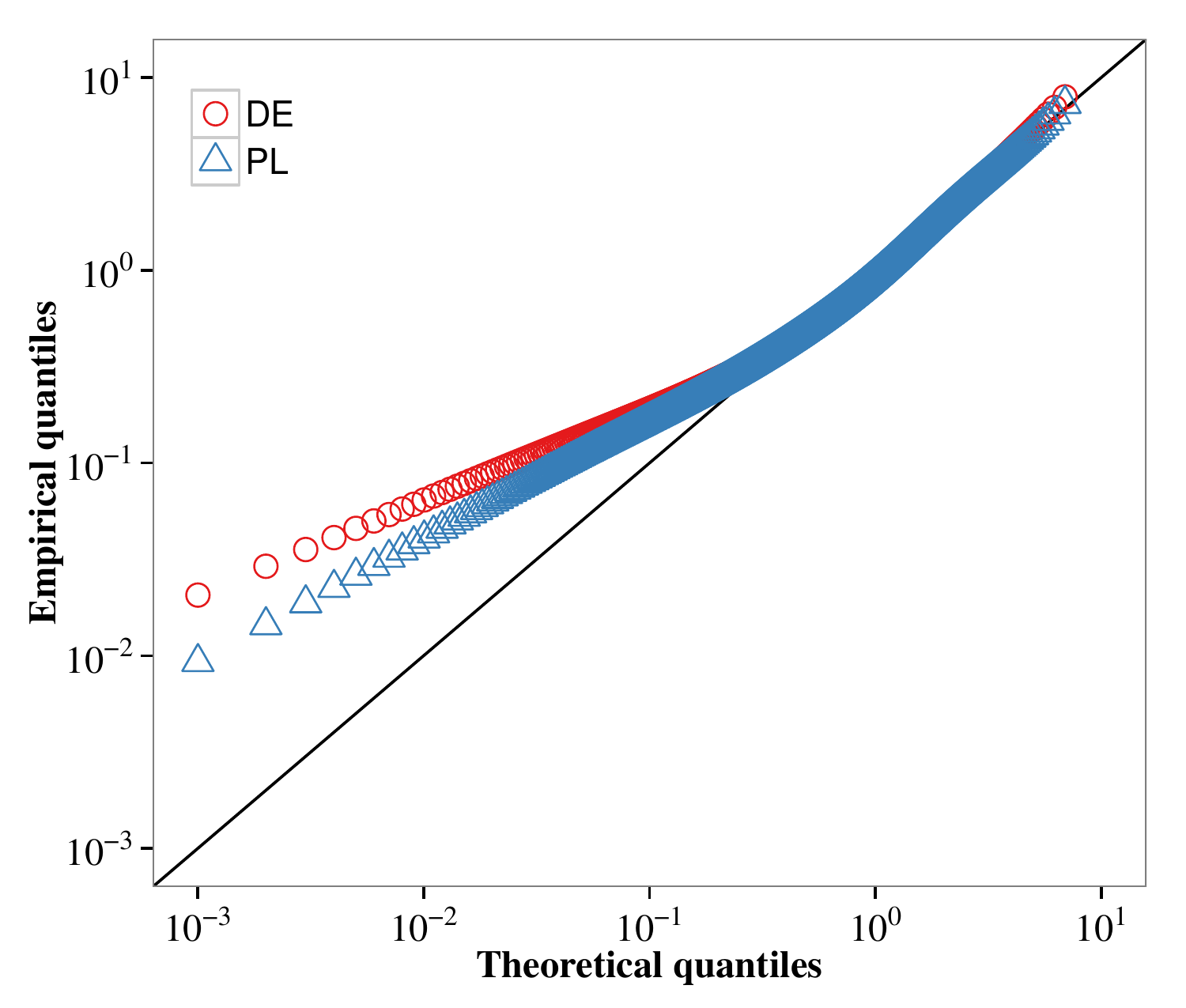}
\caption{(Color online) Comparison of the distribution of the residuals from the double exponential and power low model against the theoretical exponential distribution. Top: lin-lin scale. Bottom: log-log scale. The residuals from all the separate days and for all the three pairs are pooled together.}
\label{fig:residuals_qq}
\end{figure}

\subsection{Results}
\label{sec:results}

We estimated the best parameters for each day (i.e. the 9-hour window from 07:30 to 16:30 UTC) separately by maximizing the log-likelihood. In this way we obtain a set of optimal parameters for each day.  Appendix \ref{sec:like_appendix} describes all the details, both analytical and numerical, of the procedure we used in estimation.  

In order to compare the performance of the two kernels we studied the distribution of the residuals. In fact the time change theorem \cite{bauwens2009_review} states that if the true data generating process is a point process with intensity $\lambda$, then the residuals $$\Lambda_i=\int_{t_i}^{t_{i+1}} \lambda(s) \diff s$$ are independent and identically distributed with a standard exponential distribution. In Figure \ref{fig:residuals_qq} we compare by means of a quantile-quantile plot the empirical distribution of the residuals of both the double-exponential kernel and the power-law kernel against the standard exponential distribution. We find that the power-law kernel performs better than the double exponential. In fact, both the right tail and the left tail of the residual distribution for the power law kernel are closer to the theoretical one than the one for the exponential kernel. An analysis of the log-log plot in Figure \ref{fig:residuals_qq} indicates that the major discrepancies occur for the left tail of the distribution, which corresponds to the residuals computed on short durations. The duration distribution of the data is peaked around 100 ms even after the randomization procedure. The double exponential kernel that presents its maximum at $t=0$ is not able to reproduce this feature, while the power law kernel, which is equipped with an exponential cutoff at short lags performs better. To support the conclusions drawn from the QQ-plot we mention that we performed a test of exponentiality on the residuals, namely the Excess Dispersion (ED) test of Ref. \cite{engle&russell1998}. The test was performed on each day separately for the three currency pairs. The null hypothesis of the residuals being exponentially distributed was rejected at the 5\% level in 92.8\% of the days for the DE kernel and only in 42.2 \% of the days for the PL kernel (aggregating across the pairs).  
To conclude the comparison between the two kernels, we stress that also likelihood values are consistently higher for the power law kernel, thus favouring this choice.  

\begin{figure}[ht]
\centering
\includegraphics[width=0.5\textwidth]{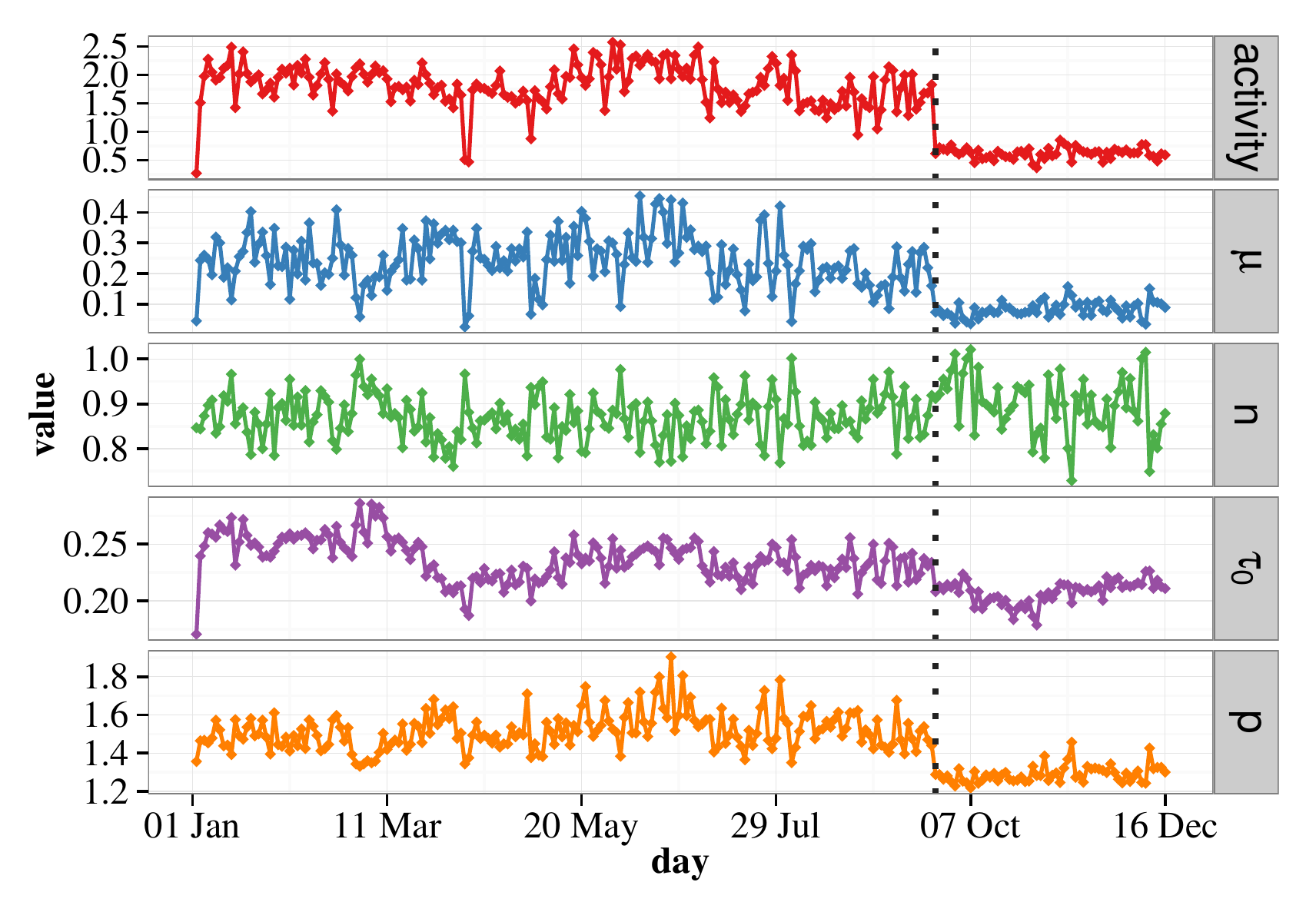}
\caption{(Color online) Time series of the Hawkes model parameters for EUR/USD. Market activity in events per second is also shown. The dotted line signals the tick size increase.}
\label{fig:trend_eurusd}
\end{figure}

We now turn to the analysis of the values obtained for the parameters. We report the results from the power law kernel. Figure \ref{fig:trend_eurusd} shows the time series of parameters estimated on EUR/USD data. Trading activity measured in events per second is also shown. In Table \ref{tab:average_bef_after} we report the average value of the  parameters obtained with the power law kernel before and after the tick increase for all the three pairs. The uncertainties on the values, as estimated from the Hessian of the log-likelihood, are about $10\%$  on $\mu$, $ 2\%$ on $n$ and $\tau_0$, and about $1\%$ on $p$.  
%
\begin{table}[h]
\centering
\begin{tabular}{lccc}
\toprule
& \multicolumn{3}{c}{Before tick increase} \\
\midrule
& EUR/USD & EUR/JPY & USD/JPY\\ 
\midrule
activity (events/s) & 1.82 (20) & 1.08 (28) & 0.76 (31)\\ 
$\mu$ ($s^{-1}$) & 0.24 (35) & 0.17 (44) & 0.12 (36)\\ 
$n$  & 0.87 (6) & 0.86 (9) & 0.84 (6) \\ 
$\tau_0$ ($s^{-1}$) & 0.24 (8) & 0.20 (11)& 0.21 (7)\\ 
$p$ & 1.51 (6) & 1.46 (10) & 1.52 (7)\\
\toprule 
& \multicolumn{3}{c}{After tick increase}\\
\midrule
& EUR/USD & EUR/JPY & USD/JPY \\
\midrule
activity (events/s)  & 0.62 (15) & 0.59 (18) & 0.22 (26) \\ 
$\mu$ ($s^{-1}$) & 0.08 (31) & 0.08 (32) & 0.03 (36)\\ 
$n$  &  0.90 (7) & 0.91 (7) & 0.95 (8) \\ 
$\tau_0$ ($s^{-1}$) & 0.21 (4) & 0.22 (16) & 0.18 (4) \\ 
$p$ & 1.29 (3) & 1.26 (3) & 1.22 (3) \\ 
\bottomrule
\end{tabular}
\caption{Average values of the parameters before and after the tick size increase. Standard deviations across the sample expressed in percent is also reported in parentheses.}
\label{tab:average_bef_after}
\end{table}

We note that the periods before and after the tick increase are very different. This is true also for the other pairs. After the tick increase the number of quote changes per second drops sharply as expected. In fact, after the increase a change in a quote implies a larger movement in the posted price.  This drop is mirrored by a drop in the baseline intensity $\mu$. The parameter $p$ controlling the power law decay decreases sharply after the tick increase, indicating an increase of the number of long durations, in accordance with what is observed in the data. The criticality parameter $n$ is quite close to $1$ and relatively insensitive to the tick size change.  


Overall, the Hawkes model equipped with the power law kernel provides a pretty good description of the empirical data. In line with \cite{bouchaud_hawkes}, we find that a large part of the activity is explained by the self-exciting mechanism ($n$ close to 1) and hence appears to be endogenously generated.

\section{Modeling market activity around macro news}\label{sec:Hawkes_news}

We now present the main original contribution of this paper, namely an extension of the Hawkes model including a news-related (exogenous) term. The model aims at reproducing in a Hawkes framework the impact of news arrival on market activity. It is important to note that the news process is considered as given, that is, the model is not meant to describe the news process itself. The news process is completely deterministic in this context. As we said, this is actually also the real case in FX market, where the news and the forecast value, when available, are announced at predetermined times.

Let $N^{\mbox{\tiny{news}}}_{t}$ be the counting process describing the arrival of a news. In order to avoid confusion with the times $\{t_i\}_{i\in \N}$ of arrival of the market activity process, we indicate with $\{z_j\}_{j\in \N}$ the arrival time of the  $j$-th news. The intensity of the market activity counting process is then
\begin{equation}
\label{eq:news_model}
\begin{split}
\lambda(t)&= \mu + \int_{-\infty}^t \phi(t-s) \diff N_s + \int_{-\infty}^t \phi_N(t-s) \diff N^{\mbox{\tiny{news}}}_s \\
&= \mu + \sum_{t_i < t} \phi(t-t_i) +\sum_{z_j < t} \phi_N(t-z_j) 
\end{split}
\end{equation}
where $\phi(t)$ is the endogenous self-exciting kernel and $\phi_N(t)$ is the exogenous kernel that accounts for the news-induced excitement. 

The condition for stationarity is obtained by taking the unconditional expectation on both side of \eqref{eq:news_model},
\begin{equation}
\label{eq:news_model_stationarity}
\begin{split}
{\Lambda}
&=\mu + \int_{-\infty}^t \phi(t-s) \E{\diff N(s)} + \int_{-\infty}^t \phi_N(t-s) \E{\diff N^{\mbox{\tiny{news}}}(s)}=\\
&=\mu + \int_{-\infty}^t \phi(t-s) \E{\lambda(s)}\diff s+ \int_{-\infty}^t \phi_N(t-s) \E{\lambda_N(s)}\diff s=\\
&=\mu + {\Lambda}\int_{-\infty}^t \phi(t-s) \diff s +{\Lambda}_N\int_{-\infty}^t \phi_N(t-s) \diff s=\\
&= \mu + {\Lambda}\int_{0}^\infty \phi(\tau) \diff \tau + {\Lambda}_N\int_{0}^\infty \phi_N(\tau) \diff \tau
\end{split}
\end{equation}  
where ${\Lambda}_N\equiv \E{\lambda_N(t)}$ is the average intensity of the news process. Solving for ${\Lambda}$, we finally get
\begin{equation}
\label{eq:news_lambda_expectation}
{\Lambda}=\frac{\mu+{\Lambda}_N\int_0^\infty \phi_N(\tau)\diff \tau}{1-\int_0^\infty \phi(\tau) \diff \tau}.
\end{equation}
Provided that ${\Lambda}_N$ exists and is finite, the condition of stationarity is thus the same as in the pure self-exciting case, namely $\int_0^\infty \phi(\tau) \diff \tau < 1$. In particular, it is not necessary that also $\int_0^\infty \phi_N(\tau) \diff \tau <1$ in order to have stationarity of market activity. As we will see below, in real data $\int_0^\infty \phi_N(\tau) \diff \tau \gg 1$.

In order to test the model of Eq. \eqref{eq:news_model} it is necessary to specify the functional form of $\phi(t)$ and $\phi_N(t)$. As we have seen above, the double exponential kernel and the quasi-power law kernel give good results for the endogenous component, hence we adopted these forms also here. For the exogenous kernel $\phi_N$ we chose a single exponential specification of the type
\begin{equation}
\phi_N(t)=\alpha_N e^{-\beta_N t}.
\end{equation}
Here, $\alpha_N$ gives the magnitude and $\beta_N$ fixes the time scale. Moreover, it seems reasonable, at least as a first approximation, that the news impact could be described by a single timescale. 

Both by using the double exponential $\kerde(t)$ and the quasi-power law kernel $\phi_{\pla}(t)$ the expression for the log-likelihood can be analytically derived. In the former case we estimate 7 parameters, while in the latter case this number is equal to 6.

We estimated and tested our model on the series of best quote changes for the three currency pairs in our database. In each day there are on average three news announcements, and we calibrated the model on a window where only one news is present. Therefore we selected those news events that are "isolated", i.e. there is no other news in a three hour window centered at the time of the news.  In this way the news event happens always at $t=5400$ s, i.e. at the median point of the window. We then estimated the parameters of the model in each window as described in Appendix A. 
Notice that it is immediate to extend our modeling to the case where more than one news is present in the investigated time interval. In the following we  focus on windows with only one news mostly for the sake of presentation clarity. 

Since we are considering only High and Medium importance news, this filtering procedure returns a total of $266$ isolated news. The news are the same for all the pairs. The average number of quote changes in each window is $17,579$ for EUR/USD, $11,199$ for EUR/JPY, and $7,322$ for USD/JPY. 


\subsection{Comparison of models with and without news}

\begin{figure}[ht]
\centering
\includegraphics[width=.35\textwidth]{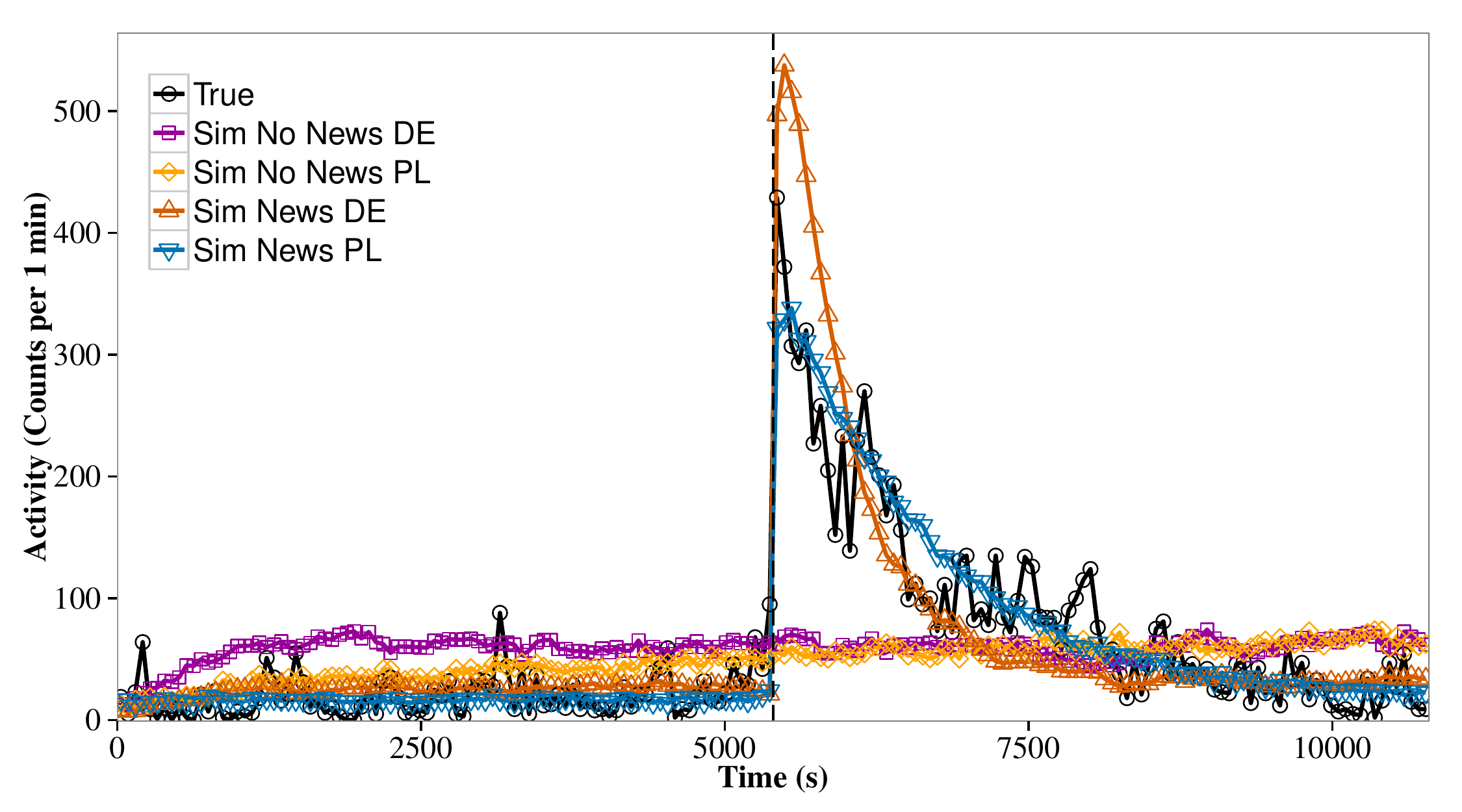}
\includegraphics[width=.35\textwidth]{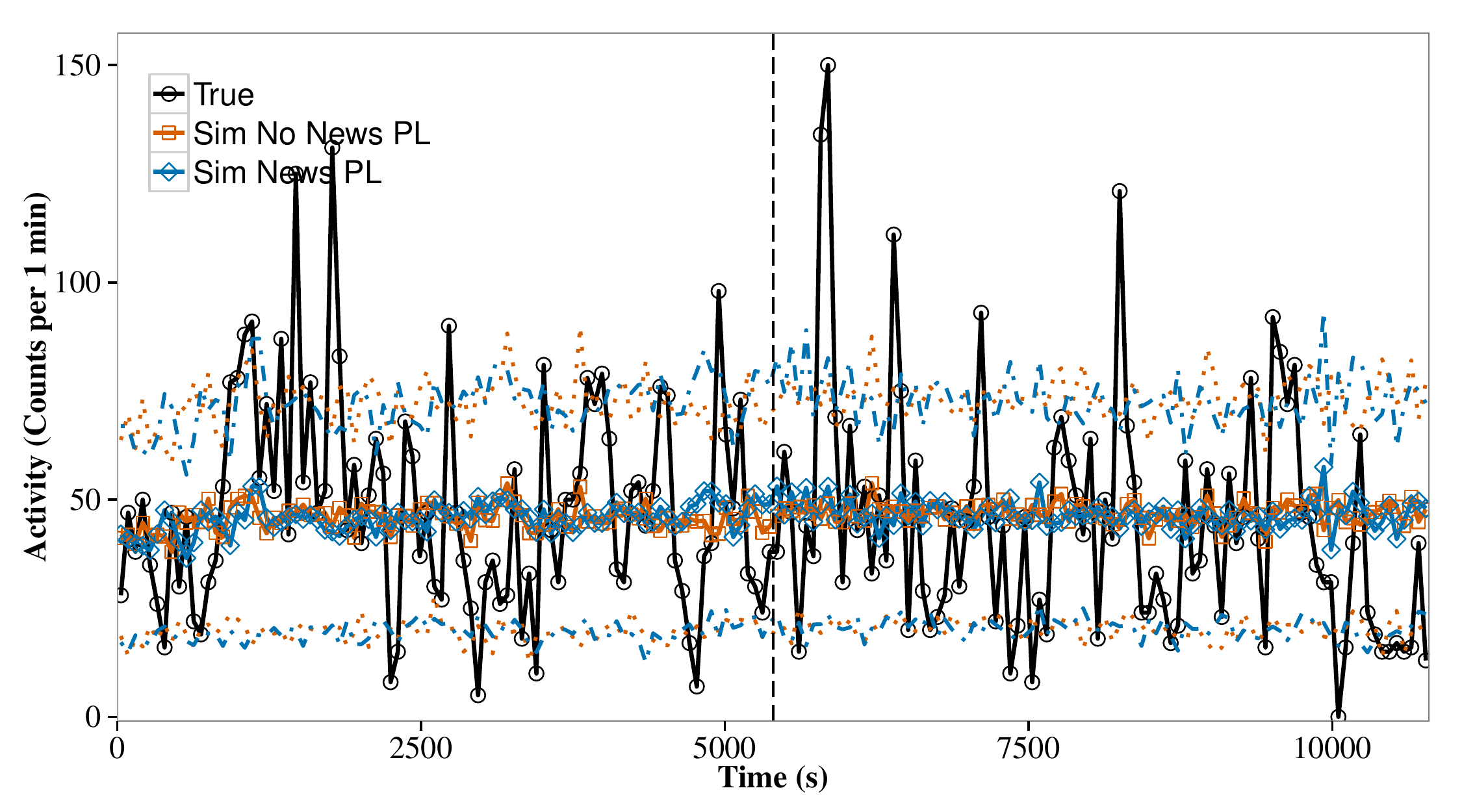}
\caption{(Color online). Actual and simulated activity around two macronews, namely  EUR/USD rate around US change in non-farm payrolls on April 6 (top panel) and USD/JPY around Euro-Zone Industrial New Orders data on January 5 (bottom panel). The dashed line corresponds to the time of the announcement. Results from simulations of the models without news are also shown for comparison. For each model we run 100 simulations and we plot the average with one standard deviation confidence bands. Confidence bands are omitted in the top panel for improving readability. Parameter estimates were $\{$ $\alpha_N^{\de}=3.2$, $\beta_N^{\de}=1.6\cdot10^{-2}$, $\alpha_N^{\pla}=1.45$, $\beta_N^{\pla}=8.2\cdot10^{-4}$ $\}$ for the first news and $\{$ $\alpha_N^{\de}=3.0$, $\beta_N^{\de}=7.0$, $\alpha_N^{\pla}=2.72$, $\beta_N^{\pla}=7.0$ $\}$ for the second one (all values in $s^{-1}$).}\label{fig:eurusd_top}
\end{figure}

In order to visualize the difference between the model with and without the news component, we present here calibrated numerical simulations of the two models and we compare them with the real data. In Figure \ref{fig:eurusd_top} (top panel) we consider an important news, namely, the release of US change in non-farm payrolls which turned out to be much lower than expected. The model without the news term fails to reproduce the trend of activity after the news. On the contrary, the model of Eq. \eqref{eq:news_model} with the endogenous power law kernel seems to reproduce pretty well both the magnitude and the temporal decay of the news impact. The model with the double exponential kernel overestimates the impact immediately after the news arrival. Moreover, it is possible to note that the models without the news term significantly overestimates the activity before the news. This is because they attribute the extra activity due to the news to the self-exciting mechanism, thus increasing the overall activity instead of concentrating it in correspondence of the news.

Figure \ref{fig:eurusd_top} (bottom panel) refers instead to the effects of the announcement of the Euro-Zone industrial new orders figures on USD/JPY. The figures were in line with expectations and the news had a scarce impact. Again, the model behaves nicely, in that it correctly captures the absence of impact in this case.

We compared the performance of the model with and without the news term across the whole set of news we examined. We simulated repeatedly both models for each news, then we examined the number of events generated by each model in the 5 minutes window that follows the news and we contrasted this number with the real number of events. The result of this comparison indicates that the model with the news term performs much better than the one without it when the news had a sizable impact. When the news produced little or no impact, the two models are essentially equivalent. Further details on this comparison are presented in Appendix \ref{sec:comparison_appendix}.

\begin{figure}
\includegraphics[width=.35\textwidth]{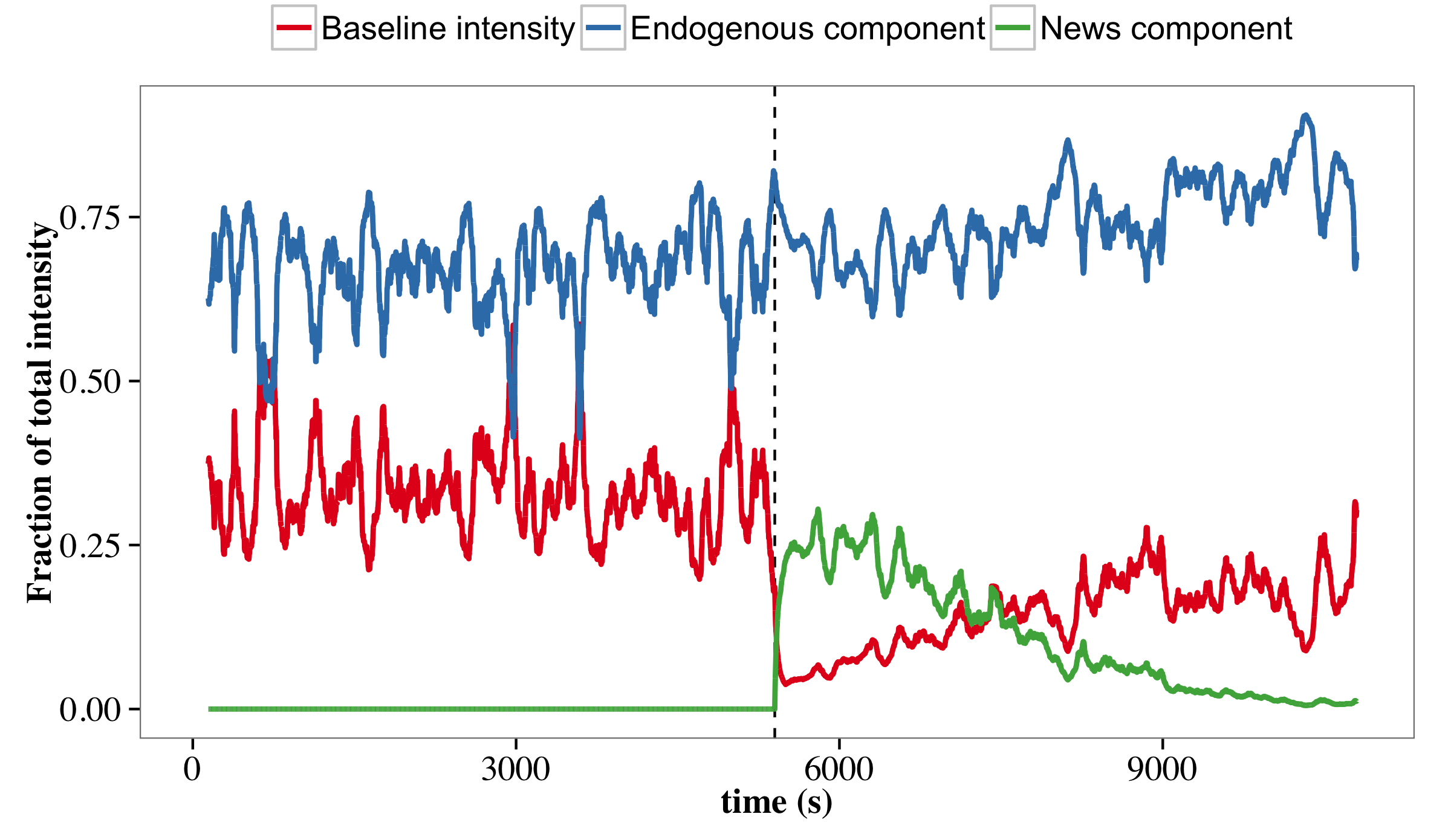}
\includegraphics[width=.35\textwidth]{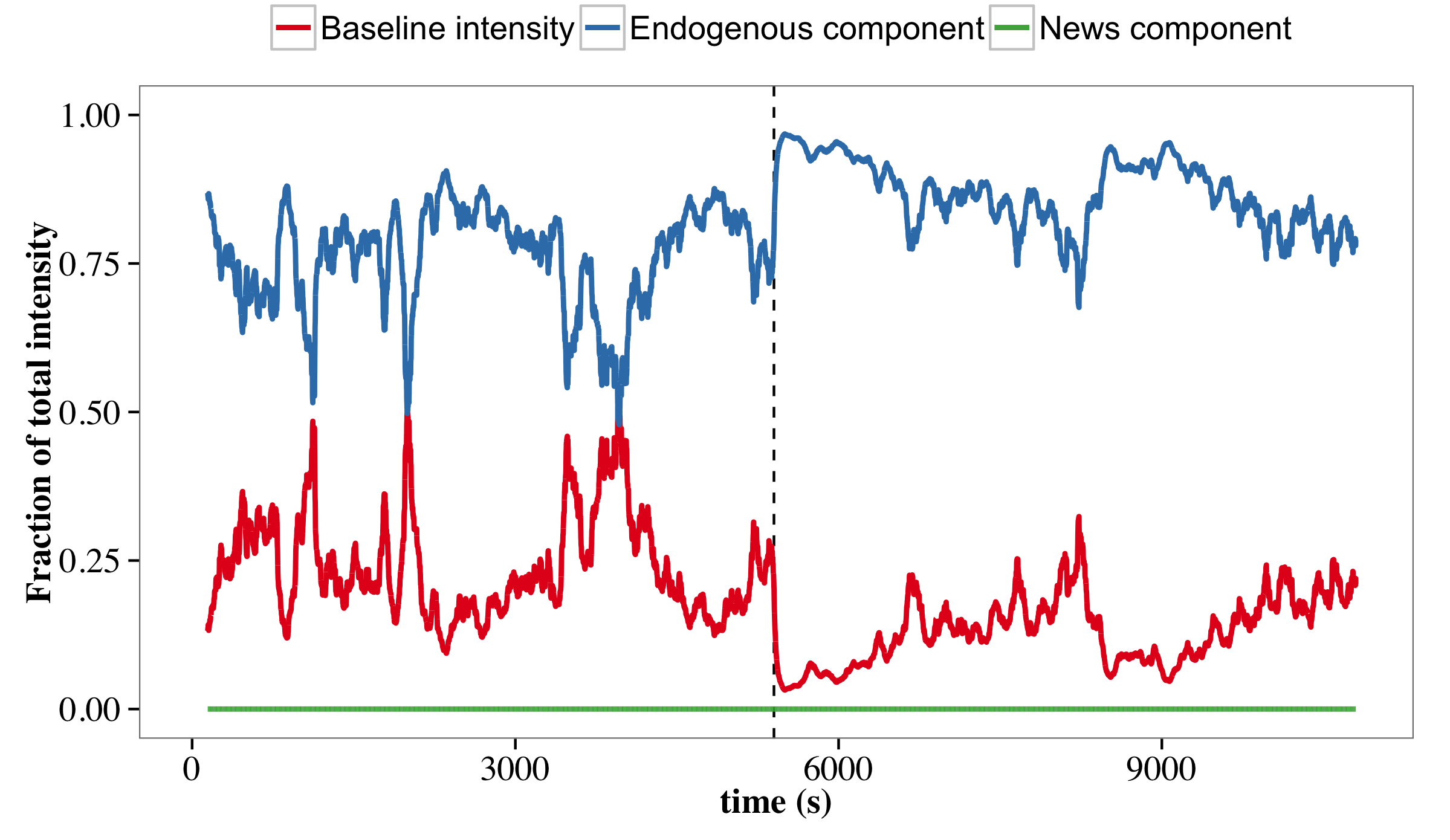}
\includegraphics[width=.35\textwidth]{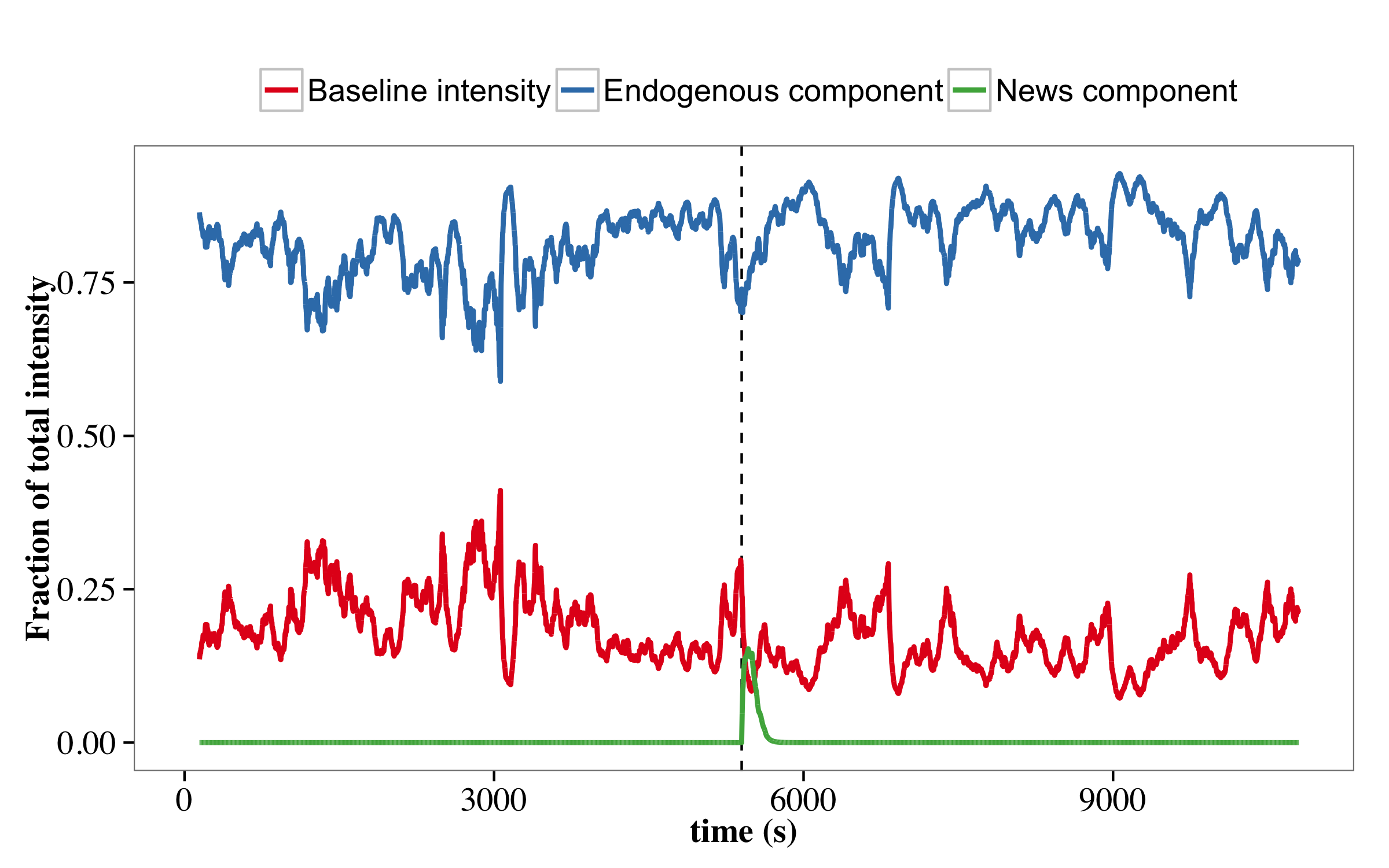}
\caption{(Color online). Components of the intensity $\lambda(t)$ as a function of time. The components sum to one and the curves are smoothed by using a simple moving average over $100\,s$. The top and middle panel refer to EUR/JPY, in correspondence of the announcement of the high importance and surprise news of US new jobs data on November 2 at 12:30. In the top (middle) panel  we consider Hawkes model with (without) the exogenous news term. The bottom panel refers to EUR/USD during a medium importance news, namely the announcement of US new home sales data on September 26 at 14:00. }
\label{fig:fractions_eurjpy_high}
\end{figure}

In our model the total intensity $\lambda(t)$ is the sum of three contributions: the baseline intensity $\mu$, the endogenous component $\int \phi(t-s) \diff N_s$, and the exogenous component $\int \phi_N(t-s) \diff N^{\mbox{\tiny{news}}}_s$. We can therefore analyse how the contribution of each term varies over time around the news announcement. The top panel of Figure  \ref{fig:fractions_eurjpy_high} shows the fraction of the intensity coming from each component as a function of time for an important news. 
After the news release, the contribution of the exogenous term rises quickly, while the weight of the endogenous one decreases. Also the contribution of the baseline intensity becomes negligible immediately after the announcement. The exogenous contribution then slowly decays towards zero and the endogenous and the baseline components regain their pre-news level. It is interesting to contrast these figures with the middle panel of Figure \ref{fig:fractions_eurjpy_high}, where the contributions to $\lambda(t)$ for the same news are shown for the model without the news term. The increment of activity after the news is now attributed to the endogenous component. The contribution of the baseline intensity has a similar trend in the two cases, though we note that its contribution before the news is higher in the model with the exogenous term. 
Finally, the bottom panel of figure \ref{fig:fractions_eurjpy_high} shows the same decomposition for a low impact news. It is possible to appreciate how, in this case, the exogenous component decays much faster and has a marginal role.

\subsection{Regression results}

The regression of the model on the real data gives a set of parameters of the endogenous and exogenous kernel for each investigated news and exchange rate.

\begin{figure}[t]
\centering
\includegraphics[width=0.28\textwidth]{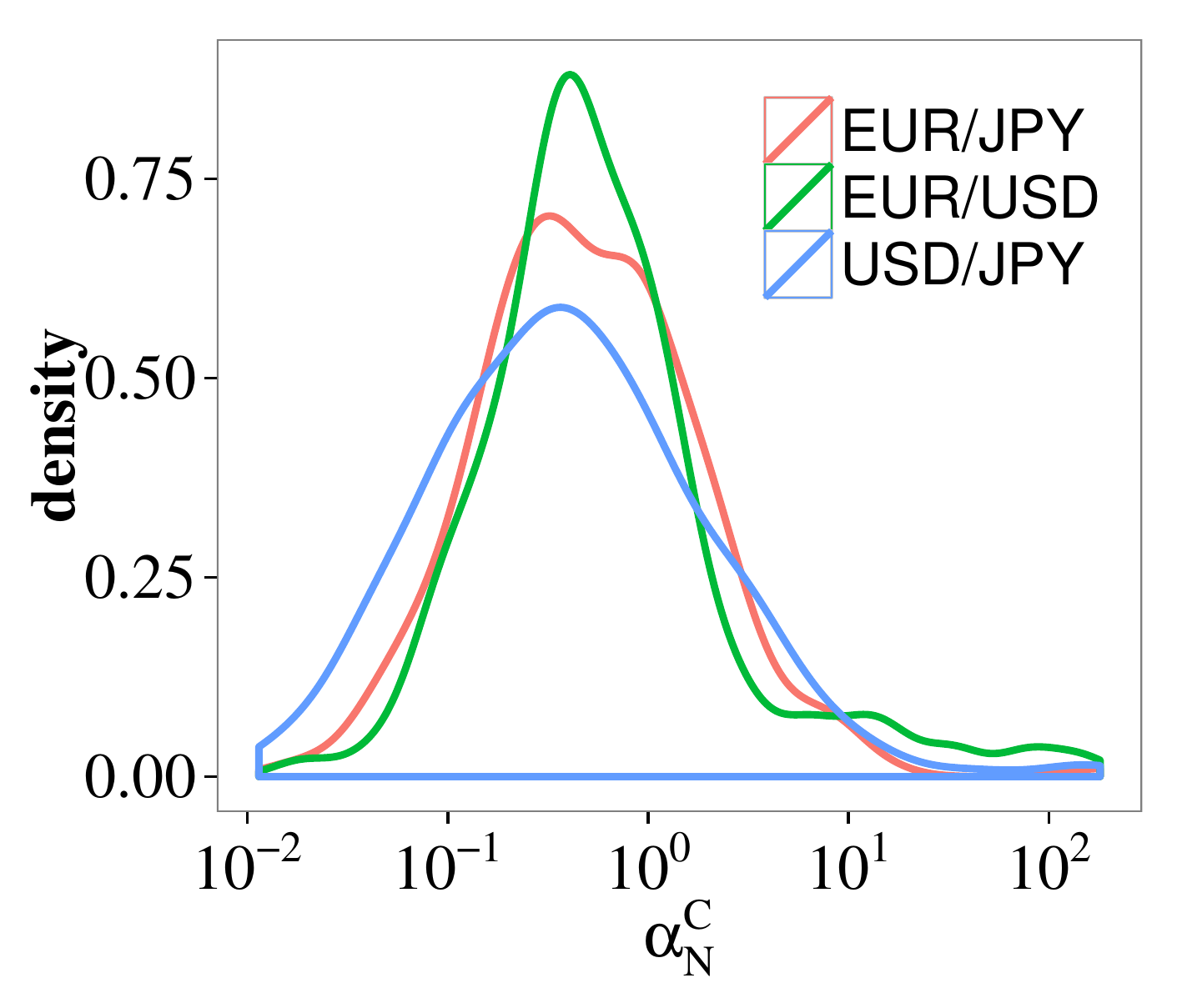}			
\includegraphics[width=0.28\textwidth]{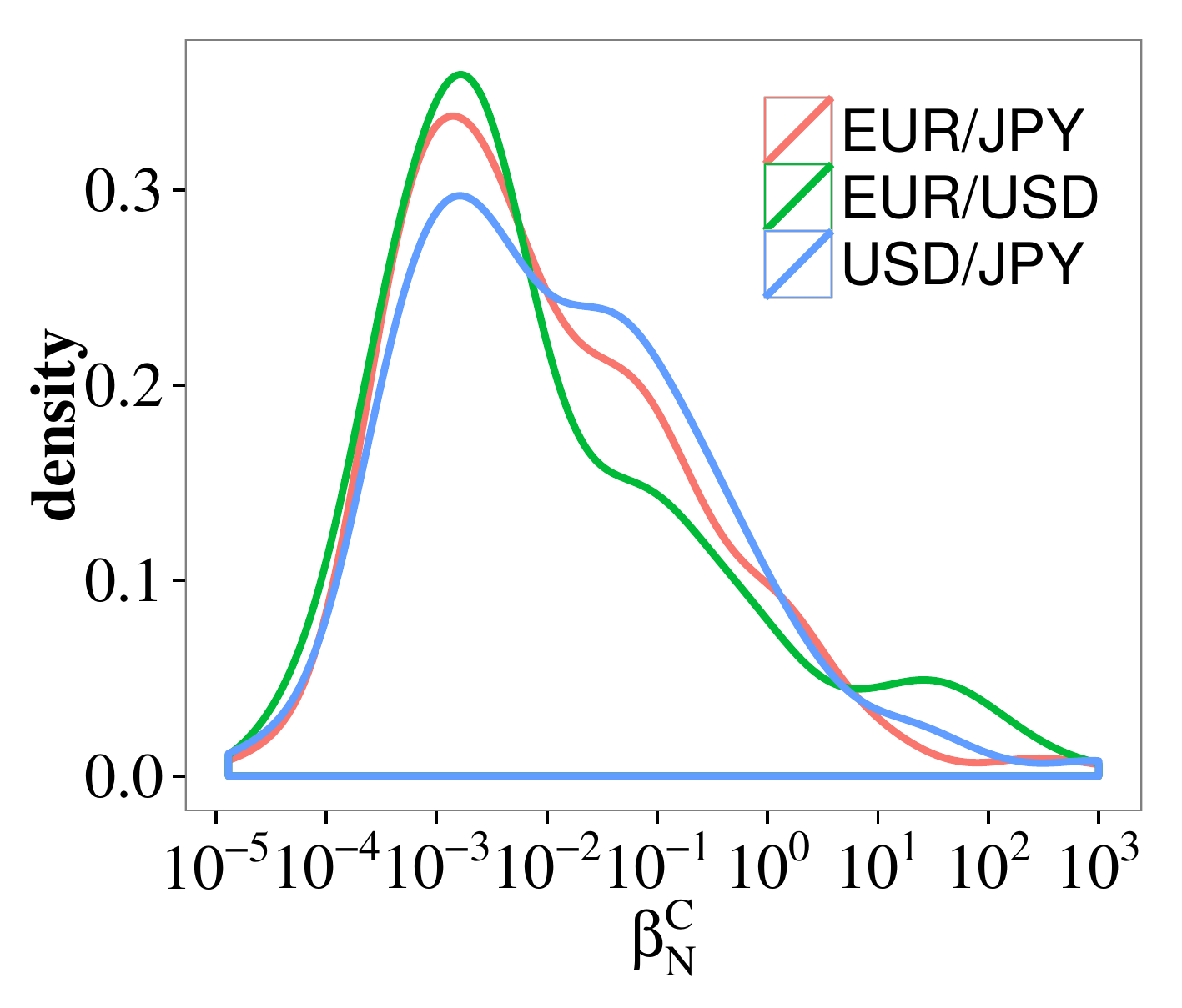}
\includegraphics[width=0.28\textwidth]{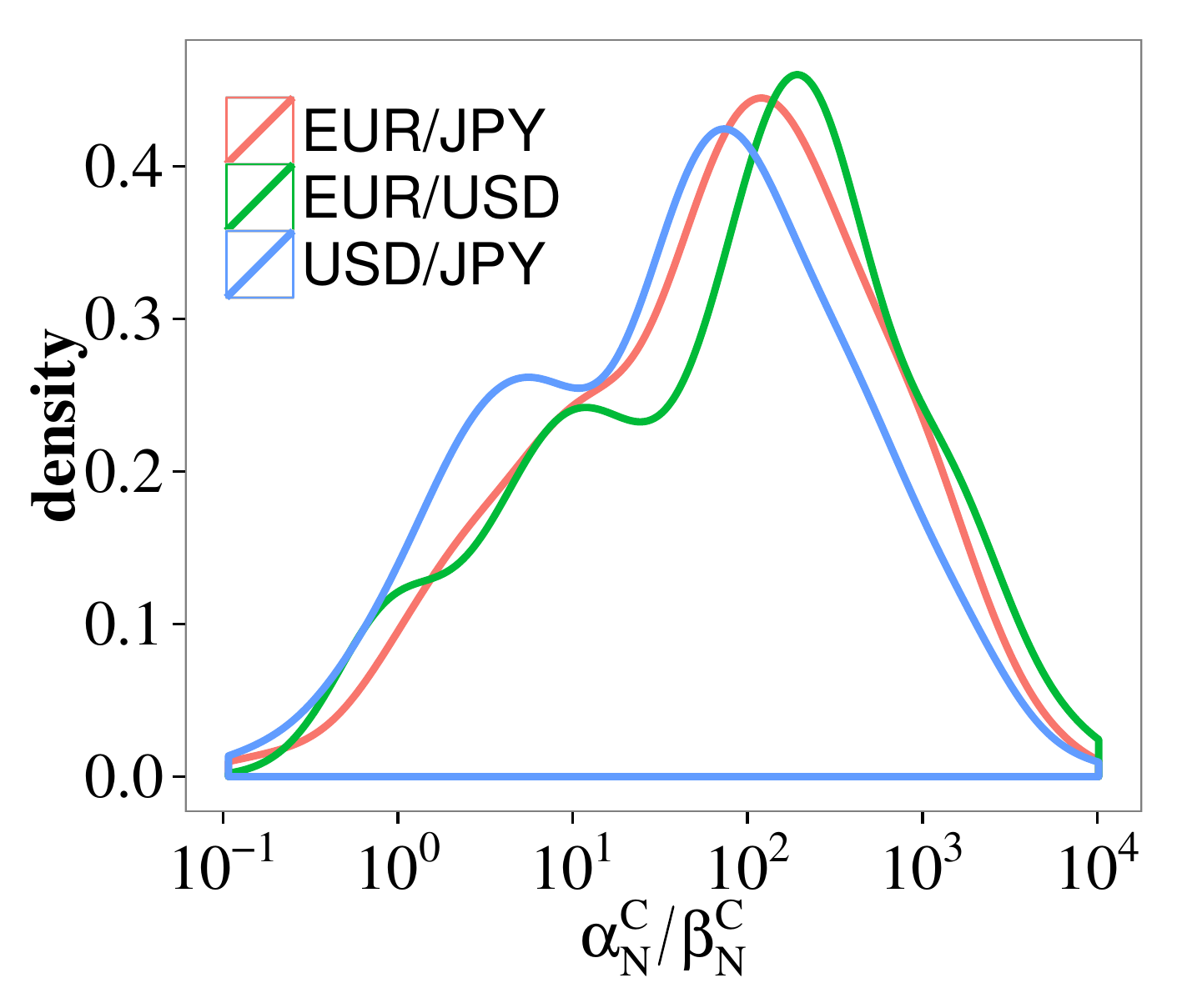}			
\caption{(Color online). Estimated probability density function of $\alpha_N$, $\beta_N$, and $\alpha_N/\beta_N$ for the three rates. 
The scale of the x axis is logarithmic, hence values of $\alpha_N$ and of the ratio $\alpha_N/\beta_N$ exactly equal to zero are not plotted. Their number is reported in Table \ref{tab:alpha=0}. }
\label{fig:dist_res}
\end{figure}

Figure \ref{fig:dist_res} shows the distribution of the parameters $\alpha_N$, $\beta_N$, and $\alpha_N/\beta_N$ for the three rates. Remarkably, the distributions are very similar across rates. 
It is worth mentioning that the figure does not show a delta peak in the distribution of $\alpha_N$ at the value zero. Their number along with the corresponding average impact is detailed in Table \ref{tab:alpha=0}. We note that the news-related parameters have a very broad distribution. This is due both to the very different effects that news produce and to the uncertainty in the estimation. As a matter of fact, uncertainty on the news-related parameters, calculated from the inverse of the Hessian matrix, is very high, with averages of about $50\%$ for both parameters. Finally, it is worth noticing that the distribution of $\alpha_N/\beta_N$ has most of its mass for values much larger than $1$.

\begin{table}
\begin{center}
\begin{tabular}{lcc}
& Number of news with $\alpha_N=0$ & $\overline{\theta}$\\
EUR/USD & 10 & 0.70\\
EUR/JPY &13 &0.69\\
USD/JPY & 13 &0.59\\
\end{tabular}
\caption{Number of $\alpha_N$ found exactly equal to zero for each pair. The average value of the corresponding $\theta$, $\overline{\theta}$, is also reported. We require $\alpha_N$ to be equal or greater then zero, and enforce this in the optimization. Hence, values equal to zero are found on the boundary of the searchable region.}
\label{tab:alpha=0}
\end{center}
\end{table}

\begin{figure}
\centering
\centering
\includegraphics[width=0.35\textwidth]{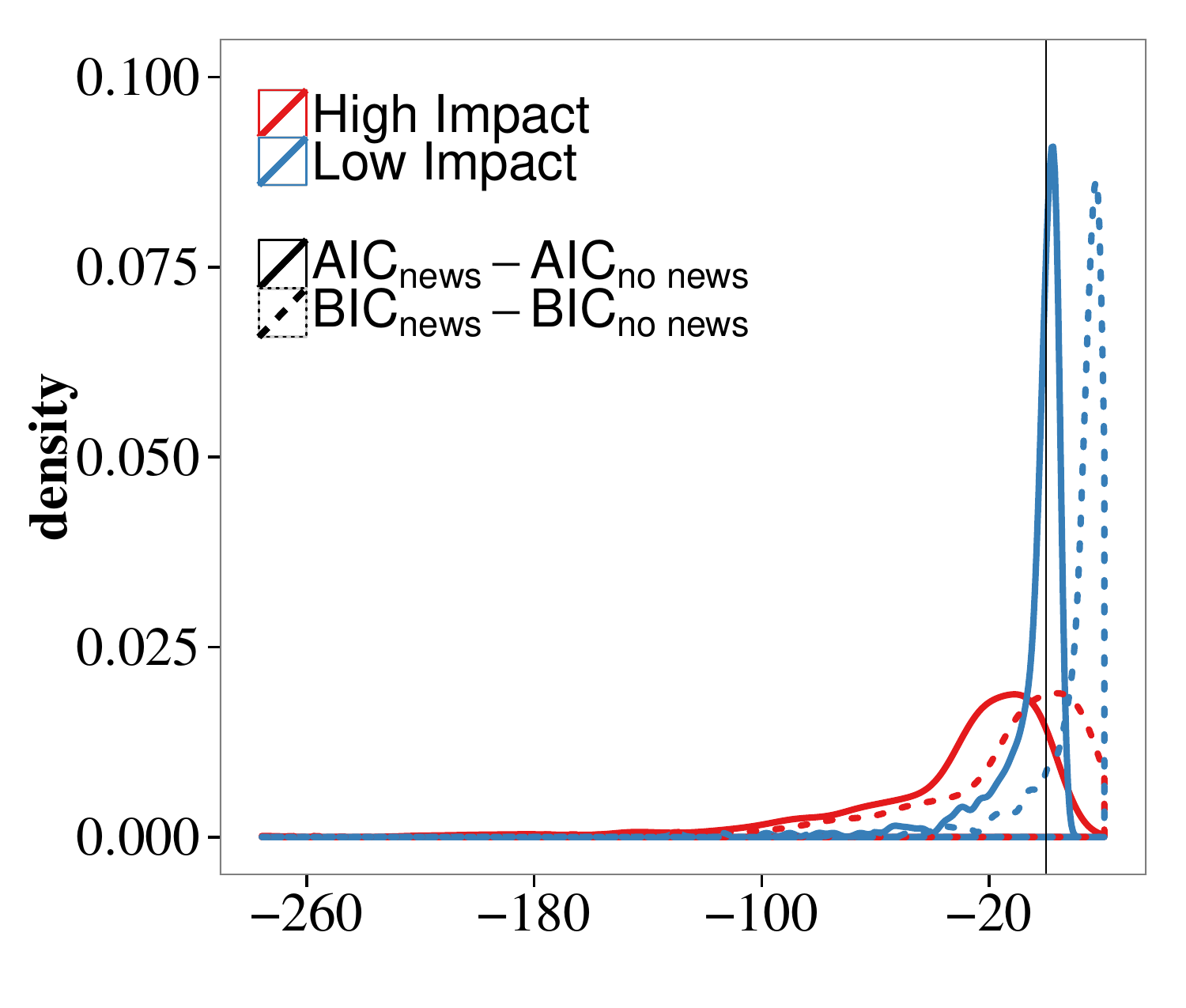}			
\centering
\includegraphics[width=0.35\textwidth]{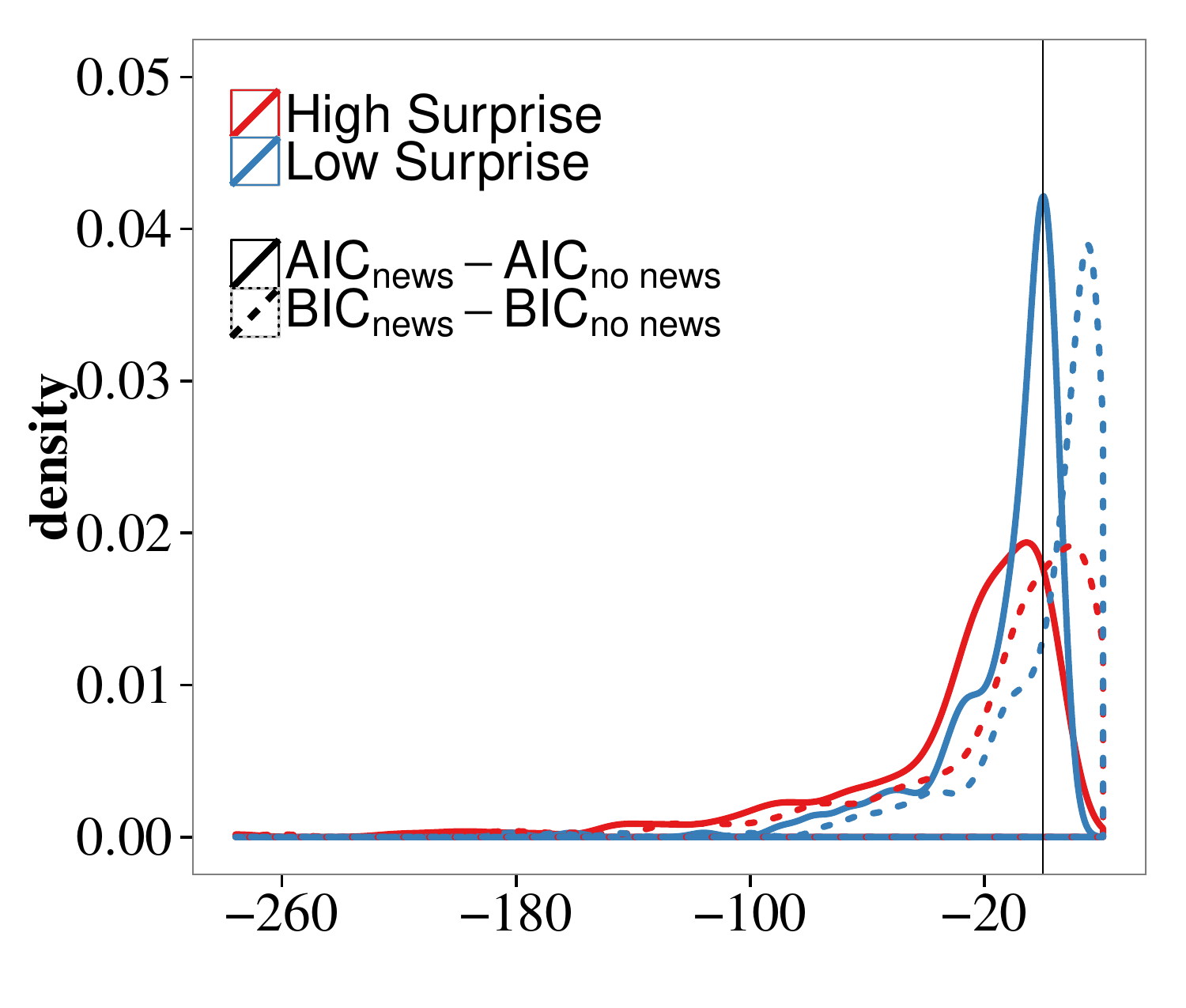}			
\caption{(Color online). Distribution of the difference in $\AIC$ and $\BIC$ scores, $\AIC_{\mbox{\tiny{News}}}-\AIC_{\mbox{\tiny{ no News}}}$ and $\BIC_{\mbox{\tiny{News}}}-\BIC_{\mbox{\tiny{ no News}}}$, between the model with the news term and the one without it. Top: distributions for high impact news, $\theta> \theta_{\mbox{\tiny{median}}}$, and low impact news, $\theta \leq \theta_{\mbox{\tiny{median}}}$. Bottom: distributions for high surprise news $S> S_{\mbox{\tiny{median}}}$, and low surprise news,  $S \leq S_{\mbox{\tiny{median}}}$. All the currency pairs are aggregated. Median values of $\theta$ for each pair are calculated before the aggregation. }\label{fig:delta_AIC}
\end{figure}

We now consider the problem of quantifying how much we improve the description of the data when we use the model with the exogenous term as compared to the performance of the simple model with only the endogenous factor. This is a typical problem of model selection \cite{Claeskens2008}. Since we can compute the maximum likelihood and the number of parameters is different in the two models, we made use of the  Akaike information Criterion (AIC) \cite{akaike1974AIC} defined as 
\begin{equation}
\AIC=2k-2\ln \like.
\end{equation}
To corroborate our results we computed also the Bayesian Information Criterion (BIC) \cite{schwarz1978}
\begin{equation}
\BIC=k\ln n -2\ln \like
\end{equation}
where $n$ represents the sample size. $\BIC$ thus generally penalizes more the introduction into the model of new free parameters than AIC does.
The model with the lowest value of $\AIC$ or $\BIC$ is the one to be selected.
Given a set of models $\{i=1,2,..,M\}$ the relative likelihood $\RL$ of model $i$ with respect to the best model in the set, for which $\AIC=\AIC_{\mbox{\tiny{min}}}$, is 
\begin{equation}
\RL=\exp{ \left( \frac{ \AIC_{\mbox{\tiny{min}}} - \AIC_i }{2}  \right)}.
\end{equation}
$\AIC$ scores for the models with the power law endogenous kernel are always better than those for the models with the double exponential kernel. This is true as well for the BIC scores. Hence, in the following we compare the results of the models with the power law endogenous kernel that differ for the presence of the news term. Figure \ref{fig:delta_AIC} shows the distribution of the difference in $\AIC$ and $\BIC$ scores between the model with the news term and the one without it, namely $\AIC_{\mbox{\tiny{News}}}-\AIC_{\mbox{\tiny{no News}}}$, and $\BIC_{\mbox{\tiny{News}}}-\BIC_{\mbox{\tiny{no News}}}$. The distribution is calculated separately for low impact and high impact news, and for high surprise and low surprise news. The figure is obtained pooling the values from all currency pairs.
For AIC, only a small fraction of the differences are positive, meaning that the model with the news term is almost always better. However, for low-impact and low surprise news, there is a sharp peak around zero. A difference in AIC scores exactly equal to 4 corresponds to the case when the two models have the same likelihood, and hence the model with more parameters is penalized by a factor two times the number of extra parameters (which is two). Hence values around 4 mean that the models are essentially equivalent.
As noted before, BIC penalizes more the extra parameters, so there is a larger portion of positive differences. However, for high-impact and high-surprise news the bulk of the distribution is on negative values. Indeed, negative differences attain even very large absolute values (as much as 260). 

Looking at the average relative likelihoods $\RL$ for the two groups we obtain
\begin{center}
\begin{tabular}{lcc}
\toprule[1pt]
Indicator&High&Low\\
\midrule
$\theta$& $5.6\cdot 10^{-8}$ & $1.0 \cdot 10^{-1}$\\
$S$ & $1.5\cdot 10^{-7}$&	$1.3\cdot 10^{-3}$\\
\bottomrule
\end{tabular}
\end{center} 
This means that for high impact news the model without the news term is about $10^{-8}$ times as probable as the model with the news term to minimize the information loss, i.e. the model with the news component is much better. Unsurprisingly, for low impact news the difference between the models reduces, albeit $\AIC$ still favours the model with the news term.

As a final investigation we consider how the endogenous term parameters are influenced by the presence of the news term. For both models with the news term and the one without it, we consider separately the distribution of the parameters for news with $\theta\leq \theta_{\mbox{\tiny{median}}}$ ("Low") and those with $\theta> \theta_{\mbox{\tiny{median}}}$ ("High").

We find that the parameter that exhibits the most relevant difference when the news term is included is $n$. For high impact news $n$ takes lower values in the model that includes the news term. In fact, an important part of the activity is now explained by the news term, whereas, in the previous model, it was attributed to the endogenous term. The baseline intensity $\mu$ has instead higher values in the model with the news term. This is at first sight somewhat surprising, since one could expect that without the news term a higher part of the observed intensity is attributed to the baseline intensity $\mu$. The introduction of the news term impacts on $n$, reducing its value when the contribution of the news is significant (i.e. when the news term explains a large part of the activity which was previously attributed to the endogenous mechanism). 
Thus, also the contribution of the endogenous component to the average intensity $\Lambda$ is lowered. 
On the other hand, the exogenous component contributes to $\Lambda$ with $\frac{\Lambda_N \cdot \frac{\alpha_N}{\beta_N}}{1-n}$. In our case, $\Lambda_N=1/5400$. 
What we observe is that this contribution is not enough to compensate for the lower values of $n$, and hence a higher $\mu$ is needed to maintain the same value of $\Lambda$. The tail of the endogenous kernel is estimated to decay faster to zero in presence of the news term and for high impact news. Instead, for low impact news, both models give about the same values. Finally, the values of $\tau_0$ are not very sensitive to the presence of the news term. 

In conclusion, the endogenous parameters are influenced by the presence of the news term. In particular, after the news an important fraction of the intensity is now attributed to the news term. We note that the value of $n$ estimated with the news term may be underestimated when no news are present (e.g. before the announcement) and this in turn leads to higher values of $\mu$ to compensate. This result is due to the fact that we employ constant parameters, while probably the "right" values of $n$ before and after a news are different. As we have seen, on average a large fraction of activity ($\approx 0.9$) is explained by the endogenous term. Our intuition is that far from news events $n\approx1$, i.e. activity is almost entirely endogenously driven, whereas after an important news $n$ becomes much smaller and the exogenous component gains weight.

\subsection{Relation between the kernel parameters and the properties of the news.}
\label{sec:news_par_surp}

In this Section we investigate how the parameters estimated by the Hawkes process with the exogenous news term are correlated with other measures of the effect of the news on the market activity, namely the jump parameter $\theta$ and the news surprise $S$ (see Section \ref{sec:news_surprise}).   

We study which parameter of the news kernel is mostly affected when a news with high impact arrives. To this end, for each rate we split the sample of news in two subsamples, the first corresponding with news with $\theta > \theta_{median}$ and the second with $\theta < \theta_{median}$. For each of the three parameters $\alpha_N$, $\beta_N$, and $\alpha_N/\beta_N$ we perform a t-test  of the hypothesis that the means for each parameter are the same for high and low impact news. The results are summarized in Table \ref{tab:t_test_theta}. We observe that, while we cannot reject the hypothesis that $\alpha_N$ and $\beta_N$ individually have the same mean in the two samples, the ratio $\alpha_N/\beta_N$ has different sample mean for large and small $\theta$. The behaviour is similar for all the rates. This means that, even when a news has a large impact, the type of response of the kernel can affect either the amplitude or the time decay of the kernel (or both), but not in a systematic way. The ratio between these quantities instead shows very different results for high and low impact news. 

\begin{table}
\begin{center}
\begin{tabular}{l|cccc}
& parameter&$t$ &$df$& $p-$value\\
\midrule
\multirow{3}{*}{EUR/USD} & $\alpha_N$ & -0.11& 238.28& 9.1$\cdot10^{-1}$ \\
& $\beta_N$ & -1.27 & 144.30 & 2.1$\cdot10^{-1}$\\
&$\frac{\alpha_N}{\beta_N}$&3.95& 181.06& 1.1$\cdot10^{-4}$ \\
\multirow{3}{*}{EUR/JPY} & $\alpha_N$ &-1.05& 133.28& 3.0$\cdot10^{-1}$\\ 
& $\beta_N$ & -1.50& 132.00& 1.35$\cdot10^{-1}$\\
&$\frac{\alpha_N}{\beta_N}$&4.75& 157.13& 4.5$\cdot10^{-6}$ \\ 
\multirow{3}{*}{USD/JPY} & $\alpha_N$ &-1.32& 133.81& 1.9$\cdot10^{-1}$ \\ 
& $\beta_N$ & -1.65& 132.00& 1.0$\cdot10^{-1}$ \\ 
&$\frac{\alpha_N}{\beta_N}$& 3.24& 166.21& 1.4$\cdot10^{-3}$\\ 
\end{tabular}
\caption{Results of the t-test of the hypothesis that the parameters for high impact and low impact news have the same mean. The test is two sided and the Welch approximation is used for the degrees of freedom ($df$) to account for unequal variances in the two samples. We report the value of the $t$-statistic along with the computed $p-$value.}
\label{tab:t_test_theta}
\end{center}
\end{table}

\begin{figure}[t]
\centering
\includegraphics[width=0.35\textwidth]{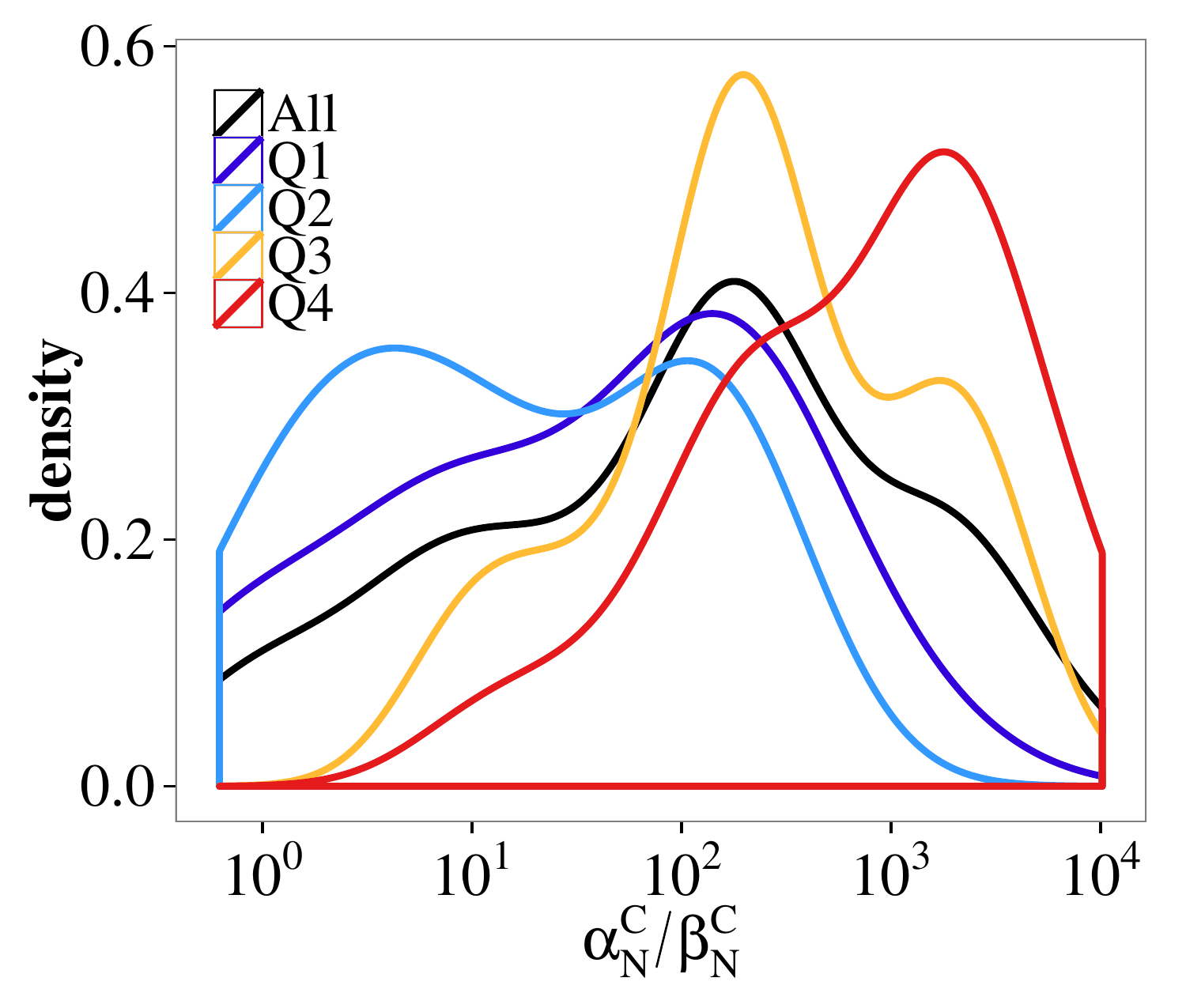}
\caption{(Color online). Estimated probability density function of the news parameter ratio $\alpha_N/\beta_N$  for four different groups formed on news surprise. Q1 indicates the fist quartiles of surprise, Q2 the second and so on. Also the total distribution is plotted for comparison. The rate is EUR/USD and we considered only High importance news.}
\label{fig:dist_ratio_surp}
\end{figure}

Finally, we examine the relation between the values of the parameters and the surprise indicator $S$ defined in Section \ref{sec:news_surprise}. To this end, we restricted the analysis to the news for which it was possible to calculate a surprise value, removing thus those for which the Forecast and Actual fields are not available. This left us with $213$ news. Since our measure of surprise is pretty rough, we divided the sample in two subsamples, one of High Importance and one of Medium Importance news. As for $\theta$, we further divided each of subsamples in two groups, one with a surprise value above and one below the sample median. For the rate EUR/USD a clear picture emerges. The rates $\alpha_N$ and $\beta_N$ are not statistically different in the two groups, both for High and for Medium importance news. On the contrary, for High Importance news the ratio  $\alpha_N/\beta_N$ is significantly larger for large surprise news as compared with small surprise news, while this effect is less evident for Medium importance news (p-values around of 10\%). To illustrate the effect on High importance news, we divided the news and the corresponding estimated parameters into four groups based on the surprise value. The intervals are delimited by the quartiles of $S$. Figure \ref{fig:dist_ratio_surp} shows the probability distribution of the ratio $\alpha_N/\beta_N$ in the four quartiles, showing that higher surprise news correspond to higher value of the ratio. Similar results are observed for the other rates. In conclusion, even with our simple measure $S$, different levels of surprise give rise to different kernel ratio distributions. 

\section{A Hawkes process with a non causal kernel}\label{sec:noncausal}

The announcement of the macroeconomic news we are considering in our analysis is known in advance to market participants. Therefore it is natural to ask whether the point process describing the market activity is affected by the news arrival {\it before} its announcement. In terms of stochastic processes this implies the presence of a non-causal kernel, in the sense that the (pre-announced) news arrival affects the intensity at earlier times.

To consider this effect, we modify the model by introducing a non-causal term in the exogenous component of the intensity. Namely, the news kernel $\phi_N(t)$ is now given by the sum of a causal and a non-causal term
\begin{equation}
\phi_N(t)=\Theta(t) \phi_N^{\C}(t)+ \Theta(-t) \phi_N^{\NC}(t)
\end{equation}
where $\Theta(t)$ denotes the Heaviside step function and we assume $\Theta(0)=0$. 
The intensity function now reads
\begin{equation}
\label{eq:news_model_NC}
\begin{split}
\lambda(t)&= \mu + \int_{-\infty}^t \phi(t-s) \diff N_s + \int_{-\infty}^{+\infty}\phi_N(t-s) \diff N^{\mbox{\tiny{news}}}_s =\\
&= \mu + \sum_{t_i < t} \phi(t-t_i) +\sum_{z_j} \phi_N(t-z_j) =\\
&=  \mu + \sum_{t_i < t} \phi(t-t_i) +\sum_{z_j<t} \phi_N^{\C}(t-z_j) +\sum_{z_j>t} \phi_N^{\NC}(t-z_j)\\
\end{split}
\end{equation}
We chose an exponential specification also for the non-causal term, so that $\phi_N(t)$ is now written as:
\begin{equation}
\label{eq:non_causal_kernel}
\phi_{N}(t)= \Theta(t) \alpha_N^{\C} e^{-\beta_N^{\C} t} +\Theta(-t) \alpha_N^{\NC} e^{\beta_N^{\NC} t}
\end{equation}
with $\alpha_N^{\C}$, $\alpha_N^{\NC} \geq 0 $, and $\beta_N^{\C}$, $\beta_N^{\NC} >0$. 

The stationarity condition now reads

\begin{equation}
\label{eq:news_lambda_expectation}
{\Lambda}=\frac{\mu+{\Lambda}_N \left[ \int_0^\infty \phi_N^{\C}(\tau)\diff \tau + \int_{-\infty}^0 \phi_N^{\NC}(\tau) \diff \tau \right]}{1-\int_0^\infty \phi(\tau) \diff \tau}.
\end{equation}
As before, provided that ${\Lambda}_N$ exists and is finite, the condition of stationarity is thus the same of the self-exciting only case, namely $\int_0^\infty \phi(\tau) \diff \tau < 1$.

\begin{figure}[h]
\centering
\includegraphics[width=.35\textwidth]{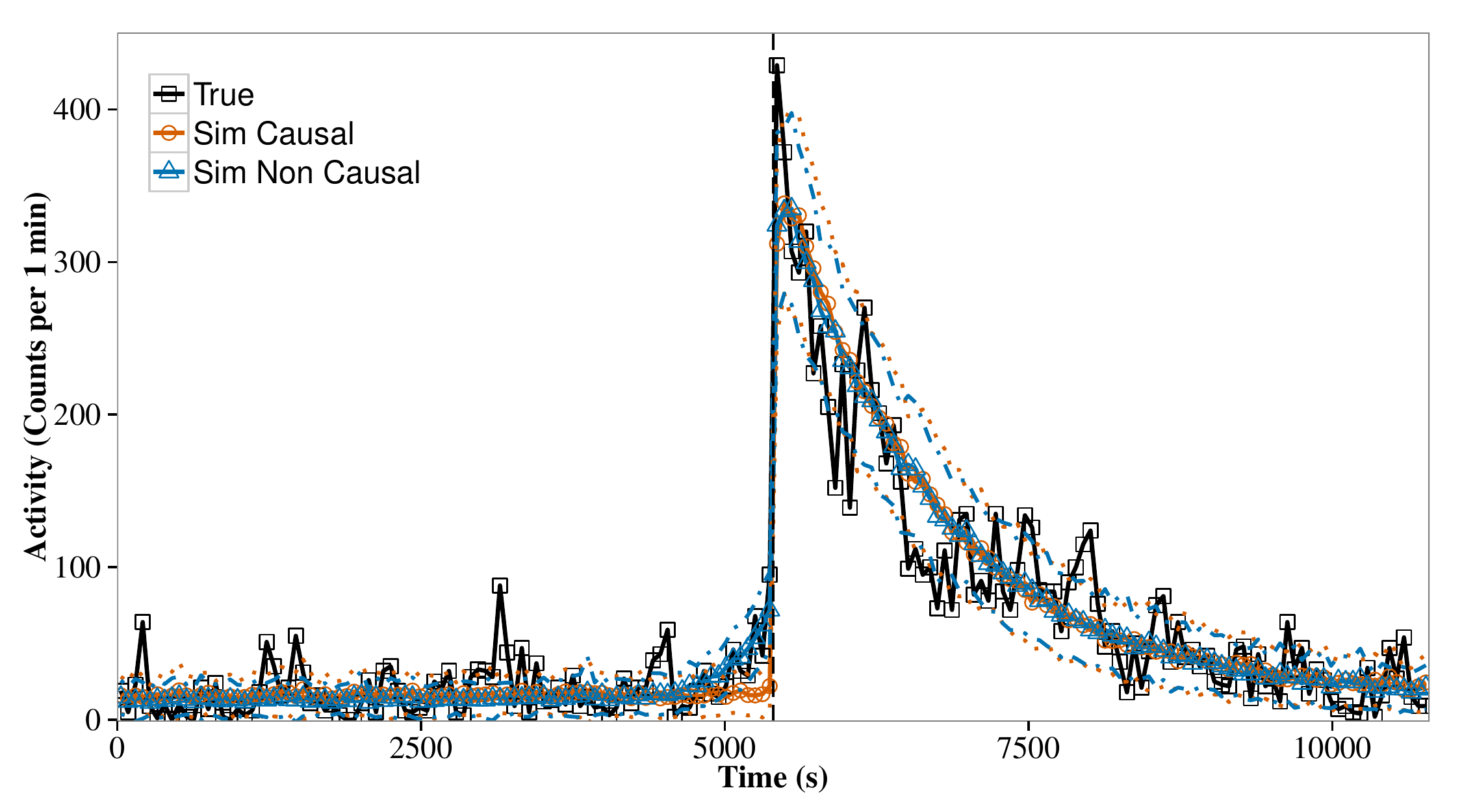}
\includegraphics[width=.35\textwidth]{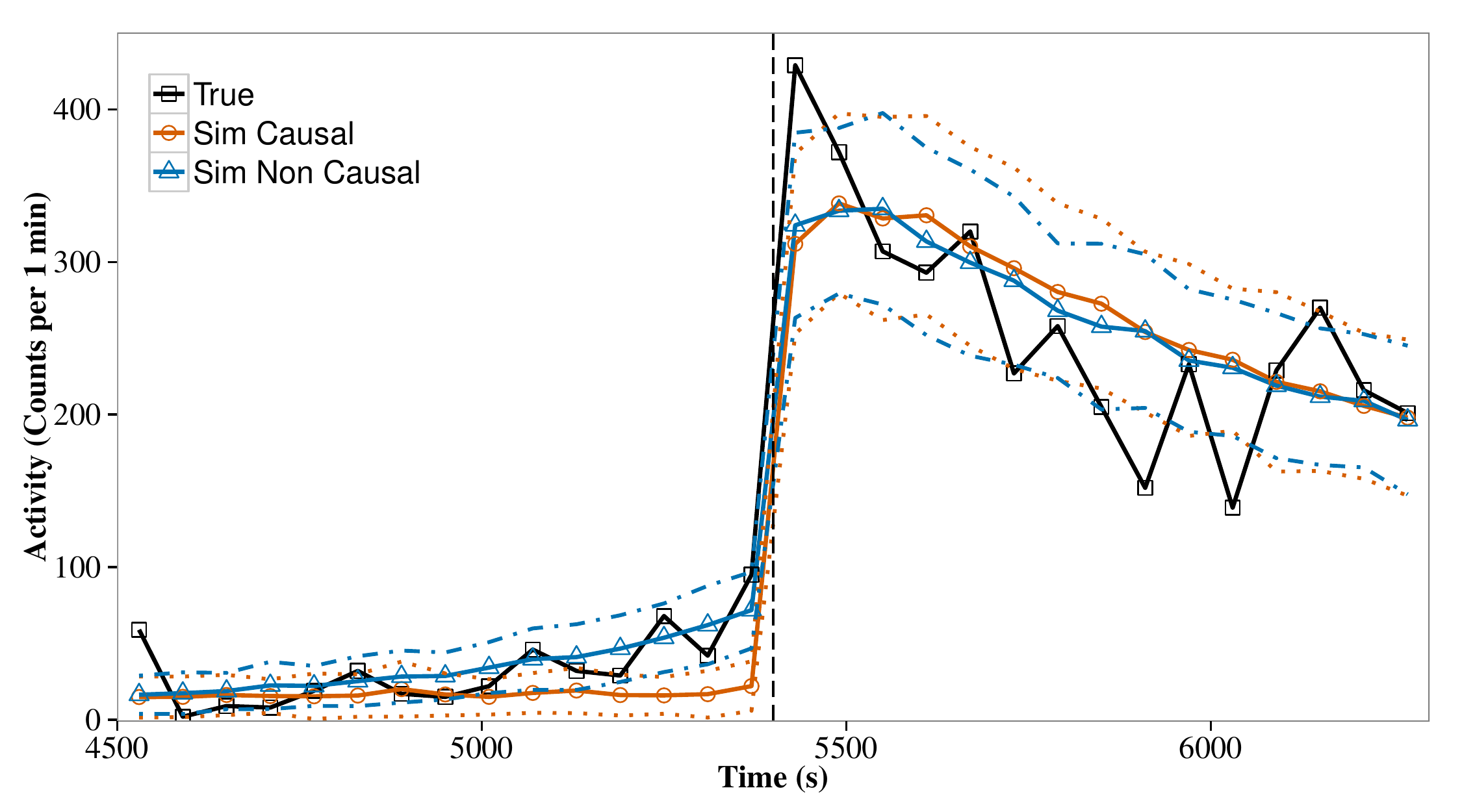}
\caption{(Color online). Actual and simulated activity measured as events per 1 min for EUR/USD on Friday April 06, when an important figure on US change in non-farm payrolls was released at 13:30. The figure produced an impact $\theta\approx 13$ and was much worse than expected (Surprise =44). The dashed vertical line corresponds to the time of the announcement.
The bottom panel is a zoom of the top panel in a 30 minute window around the news. Results from simulation of the models with and without the non causal term are compared.  The dotted and dashed-dotted lines are one standard deviation confidence intervals obtained from simulations of the calibrated model of the causal-only and non-causal model, respectively. Parameters estimates for the non causal term are $\alpha_N^{\NC}=0.3$, $\beta_N^{\NC}=3.0\cdot10^{-3}$ (all values in $s^{-1}$).  }
\label{fig:eurusd_81_C_vs_NC}
\end{figure}

We re-estimated the parameters of the model with the non-causal term on the same dataset used for the causal-only model. In Figure \ref{fig:eurusd_81_C_vs_NC}
we show simulations of the causal-only and non-causal models compared against the real data for an important news. We also add one standard deviation confidence intervals obtained by performing a large number of simulations of the model. We note that, especially by looking at the zoomed time series (bottom panel), before the announcement the confidence bands of the two models almost do not overlap, and the non-causal model reproduces quite well the news-anticipation effect when it is present, whereas the causal-only model misses completely this feature of the data. 

\begin{figure}[h]
\centering
\includegraphics[width=.35\textwidth]{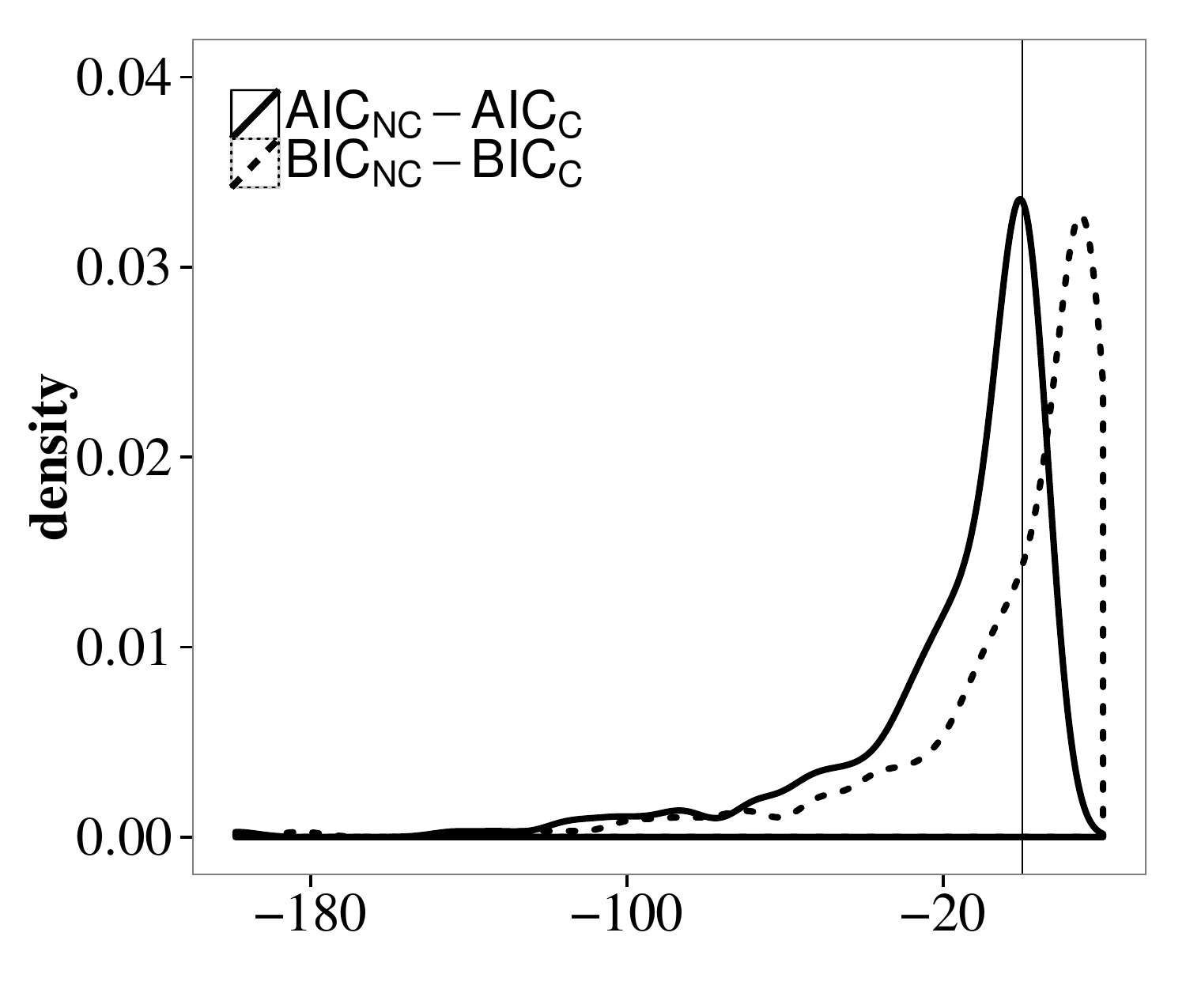}
\caption{Estimated probability density function of the differences in AIC and BIC scores between the extended model with the non-causal term and the causal-only model. Differences from all three currency pairs are aggregated.}
\label{fig:eurusd_AIC_C_vs_NC}
\end{figure}

We then compare the extended model with the causal-only one via AIC and BIC. Figure \ref{fig:eurusd_AIC_C_vs_NC} shows the distribution of the differences $\AIC_{\NC}-\AIC_{\C} $ and  $\BIC_{\NC}-\BIC_{\C} $. Results from all currency pairs are pooled together. As the relatively large number of news for which $\alpha_N^{\NC}=0$ suggests, the news-anticipation effect is often very small or absent. In these cases, the extended model with the non-causal term reduces in practice to the causal-only model. Hence, the extra parameters do not carry any benefit in these situation, and this is reflected in the information criteria scores. On the other hand, when the news anticipation effect is relevant, the improvement of the extended model is significant and AIC and BIC scores strongly favour the introduction of the non-causal term. Moreover the parameters of the causal news kernel are essentially unchanged by the introduction of the non-causal kernel.

Overall it seems that adding the non-causal term results in an improved description of the real data, albeit in some cases we simply recover the performance of the already discussed causal-only model.

\section{Conclusions}\label{sec:conclusions}

In this paper we have introduced a Hawkes process approach to the description of foreign exchange rate dynamics around the announcement of macroeconomic news. These news are very important and might affect in a significant way the trading activity in a time window around the announcement. We have first considered the unconditional Hawkes modeling, i.e. we have neglected the role of news. We found that, as seen in other asset classes, foreign exchange markets are very close to criticality, as measured by the integral of the kernel $n$.  Moreover, while the parameters of the kernel show an abrupt change around the change of the tick size, the criticality parameter $n$ seems to be pretty insensitive to it. We then considered our main original contribution, namely the introduction of a kernel describing the effect of macro news on the trading activity. Since announcements are known in advance, we have also considered the possible presence of a non causal kernel, describing how the market prepares itself to the news. In both cases we have shown that the model with the news kernel(s) outperforms the simple Hawkes model with only endogenous kernel. We also noted that, once the news term is introduced, the estimate of the criticality parameter $n$ is smaller. This suggests that in the presence of localized exogenous excitations a Hawkes model which does not consider news-triggered non stationarity could overestimate $n$. Moreover, even if the non-causal part only helps in a fraction of cases, from the practical point of view of a market maker (or anyone trading around news) being able to gauge the market activity and liquidity at those particular points is very important and there is a premium to be able to get that estimate right (or big cost to getting it wrong). Finally, we have also explored the role of the surprise of the news with respect to forecasted value on the parameters of the kernel.  

An important limitation of our model is that, as it was presented, it treats equally all the news. If more than one news is present in the estimation interval and the news have different characteristics that lead to very different impacts, the model is expected to average the differences or to be influenced by the most relevant ones. Future developments can address this issue for example by introducing a surprise-dependent amplitude $\alpha_N=\alpha_N(S)$. Limiting the dependence on surprise to the amplitude allows to conserve the recursion relation that speeds up likelihood computation. However, if the number of exogenous events is limited, one can also let $\beta_N$ depending on surprise. Finally, when the number of news is small, one could simply add a separate exogenous term for every news. This increases the number of parameters, but eliminates the need of surprise dependence. 

Despite being focused on financial market data, the approach presented in this paper could be useful in the analysis and modeling of other complex systems. In fact the presence of exogenous and endogenous drivers of the activity is ubiquitous in other systems monitored in continuous time. Describing the different source of excitations as correlated  point processes is in general quite complicated, and the Hawkes approach proposed here could be a useful method to model, fit, and evaluate the relative importance the two drivers of a generic stochastic dynamics.

\section*{Acknowledgments} 
Authors acknowledge eRisk HSBC team for support and EBS for having provided the data. FL acknowledges partial support by the grant, SNS13LILLB Systemic risk in financial markets across time scales. We also thank E. Bacry and J-P. Bouchaud for useful discussions.

\appendix
\section{Likelihood details} 
\label{sec:like_appendix}
In this appendix we give the details of log-likelihood computations.
For point process models that are specified via the conditional intensity, the likelihood function has a fairly simple expression. This makes maximum likelihood inference a popular choice for these models.

Here we follow the notation of \citet{bowsher2002}. Suppose we are observing a point process in the interval $[0,T]$. We denote $\lambda^*(t)= \lambda(t|{\hist}_{\left(-\infty, t\right] } )$ the \textit{complete} conditional intensity of the process, that is conditioned on the whole past history. We indicate with $\lambda(t)= \lambda(t|{\hist}_{(0, t]})$ the intensity conditional on the sole information from the observation interval $[0,T]$. Then, from \citet{ogata1978}, the exact log-likelihood on the interval $[0,T]$ for a simple univariate stationary point process with intensity $\lambda(t)$ is 

\begin{equation}
\label{eq:like_pp_univ}
\ln{{\like} \left( \theta ; N(t)\right)}=-\int_0^T(\lambda_{\bm{\theta}}(s)) \diff s+ \int_0^T \ln{\lambda_{\bm{\theta}}(s)} \diff   N(s),
\end{equation}
where $\bm{\theta}$ denotes the parameters vector. \citet{ogata1978} defines also a theoretical log-likelihood under the information from the infinite past
\begin{equation}
\label{eq:like_theo}
\ln{{\like}^* \left(\bm{\theta} ; N(t)\right)}=-\int_0^T(\lambda^*_{\bm{\theta}}(s)) \diff s+ \int_0^T \ln{\lambda^*_{\bm{\theta}}(s)} \diff   N(s).
\end{equation}

\citet{ogata1978} proves, under assumptions that are verified for Hawkes processes, some proprieties of the maximum likelihood estimator $\hat{\theta}^T$ obtained by maximizing \eqref{eq:like_pp_univ} with respect to $\theta$. In particular, $\hat{\theta}^T$ is found to be:
\begin{itemize}

\item \textit{Consistent}, i.e. $\hat{\theta}^T$ converges to the true value $\theta_0$ as $T \to \infty$:
\begin{equation}
\forall \epsilon >0, \;\;\; \lim_{T \to \infty} P\left[ \abs{\hat{\theta^T}-\theta_0}
>\epsilon \right]=0
\end{equation}

\item \textit{Asymptotically normal}, i.e.:
\begin{equation}
\label{eq:asympt_norm}
\sqrt{T}(\hat{\theta}^T-\theta_0) \to {\ndist} (0,I^{-1}(\theta_0))
\end{equation}
with
\begin{equation}
\label{eq:fisher_matrix}
 I^{-1}(\theta)= \E{\frac{1}{\lambda^*_{\theta_0}(t)}\dprima{\lambda^*_{\theta_0}(t)}{\theta_i} \dprima{\lambda^*_{\theta_0}(t)}{\theta_j}}=-\E{\frac{1}{T} \dparsec{\ln {\like}^*_T(\theta_0)}{\theta_i}{\theta_j } }
\end{equation}
\item \textit{Asymptotically efficient}, the following asymptotic relation exists between the Hessian of the likelihood function and the information matrix $I(\theta_0)$:  
\begin{equation}
\label{eq:hessian_like_asympt}
- \E{\frac{1}{T} \dparsec{\ln\like_T(\theta_0)}{\theta_i}{\theta_j}} \to I_{ij}(\theta_0)
\end{equation}
so the variance of the estimator reaches asymptotically the lower bound $I(\theta_0)^{-1}$.

\end{itemize}
 
These are essentially the standard asymptotic properties of maximum likelihood estimators, however there is an important difference: it is the complete intensity $\lambda^*(t)$ that enters \eqref{eq:fisher_matrix}, and not $\lambda(t)$. Hence is not true in general that $$I_{ij}(\theta)=-\E{\frac{1}{T} \dparsec{\ln\like_T(\theta_0)}{\theta_i}{\theta_j}},$$ albeit the convergence result \eqref{eq:hessian_like_asympt} supports the use of the Hessian matrix $$\frac{1}{T} \dparsec{\ln\like_T(\hat{\theta})}{\theta_i}{\theta_j}$$ to estimate $I_{ij}(\theta_0)$ \citep{bowsher2002}. 


For a linear Hawkes process in the form \eqref{eq:hawkes_def} with kernel $\phi(t)$, denoting with $\tilde{f}$ an antiderivative of a function $f$, we obtain for the log-likelihood in the interval $[0,T]$
\begin{equation}
\ln\like = -  \mu T - \sum_{t_i} \left( \tilde{\phi}(T-t_i) -  \tilde{\phi} (0) \right) + \sum_{t_i} \ln \lambda(t_i)
\end{equation}
The last term on the right hand side hides a double summation that makes the computational complexity of the expression grow as ${\cal{O}}(N^2)$, where $N$ is the number of events in the interval. This constitutes an obstacle to applications on large samples. However, for the particular case of a linear Hawkes process with kernel given by a sum of $P$ exponentials of the form
\begin{equation}
\phi(t)=\sum_{q=1}^P \alpha_q e^{-\beta_q t},
\end{equation}
\citet{ogata1981} shown that the log likelihood can be computed recursively reducing computational complexity to ${\cal{O}}(N)$. In fact, we have
\begin{align}
\label{eq:like_haw1}
\ln\like = &- \mu T - \sum_{q=1}^P \sum_{t_i} \frac{\alpha_q}{\beta_q} \left( 1-e^{-\beta_q(T-t_i)} \right)\nonumber\\
&+\sum_{i=1}^N \ln{\left(  \mu(t_i) + \sum_{j=1}^P \sum_{k=1}^{i-1} \alpha_j e^{-\beta_j(t_i-t_k)} \right)},
\end{align}
where $N$ denotes the number of events in the interval $[0,T]$.
Now, setting
\begin{align}
\label{eq:R_def}
R_j(i) &=\sum_{k=1}^{i-1} e^{-\beta_j(t_i-t_k)} \nonumber\\
&= e^{-\beta_j(t_i-t_{i-1})} \sum_{k=1}^{i-1} e^{-\beta_j(t_{i-1}-t_k)} \nonumber \\
&= e^{-\beta_j(t_i-t_{i-1})} \left( 1+ \sum_{k=1}^{i-2} e^{-\beta_j(t_{i-1}-t_k)} \right) \nonumber \\
&= e^{-\beta_j(t_i-t_{i-1})} \left(1+R_j(i-1) \right)
\end{align} 
the log likelihood expression \eqref{eq:like_haw1} can be rewritten as:
\begin{align}
\label{eq:like_haw_exp}
\ln\like  = &-  \mu T - \sum_{q=1}^P \sum_{t_i} \frac{\alpha_q}{\beta_q} \left( 1-e^{-\beta_q(T-t_i)} \right)\nonumber\\
&+ \sum_{i=1}^N \ln{\left(\mu(t_i)+ \sum_{j=1}^P \alpha_j R_j(i) \right)} 
\end{align}
with $R_j(1)=0$, $\forall j$. 

From equation \eqref{eq:like_haw_exp} it follows that for the double exponential kernel \eqref{eq:kernel_uni_de} discussed in section \ref{sec:Hawkes} we have:
\begin{align}
\label{eq:like_double_exp}
\ln\like = &- \mu T - \sum_{q=A,B} \sum_{t_i} \frac{\alpha_q}{\beta_q} \left( 1-e^{-\beta_q(T-t_i)} \right) \nonumber\\
&+ \sum_{t_i} \ln \left( \mu + \alpha_p R_q(i) \right).
\end{align}
Likewise, for the approximate power-law kernel \eqref{eq:bouchaud_kernel} the corresponding log-likelihood reads
\begin{align}
\label{eq:like_bouchaud_kernel}
\ln\like = &- \mu T \nonumber \\
&- \sum_{t_i} \frac{n}{Z} \left( \sum_{k=0}^M (a_k)^{1-p} (1-e^{-\frac{T-t_i}{a_k}}) - \frac{S}{a_{-1}} (1-e^{-\frac{T-t_i}{a_{-1}}} ) \right) \nonumber \\
&+ \sum_{t_i} \ln\left[ \mu + \frac{n}{Z} \left( \sum_{k=0}^M (a_k)^{-p} R_k(i) - S\,R_S(i) \right) \right],
\end{align}
with 
\begin{align}
\label{eq:recursive_relations}
R_k(i) &= \sum_{t_j <t_i} e^{-\frac{t_i-t_j}{a_k}}=e^{-\frac{t_i-t_{i-1}}{a_k}} \left( 1+ R_k(i-1) \right) \\
R_S(i) &= \sum_{t_j <t_i} e^{-\frac{t_i-t_j}{a_{-1}}}=e^{-\frac{t_i-t_{i-1}}{a_{-1}}} \left( 1+ R_S(i-1) \right).
\end{align}

Using equation \eqref{eq:like_pp_univ} we obtain also the expression of the log-likelihood when exogenous terms, both causal and non-causal, are added:
\begin{align}
\ln\like = &-  \mu T - \sum_{t_i} \left( \tilde{\phi}(T-t_i) - \tilde{\phi}(0) \right) \nonumber\\
&- \sum_{z_j} \left( \tilde{\phi}^{\C}(T-z_j) - \tilde{\phi}^{\C}(0) \right) \nonumber\\
&- \sum_{z_j} \left( \tilde{\phi}^{\NC}(0) - \tilde{\phi}^{\NC}(-z_j) \right)  \nonumber\\
&+ \sum_{t_i} \ln \lambda(t_i)
\end{align}
$z_j$ being the times at which exogenous events take place.

Thus, for our non-causal model \eqref{eq:news_model_NC}, the expression of the log-likelihood derived from the previous relation reads
\begin{align}
\label{eq:like_causal_non_causal}
\ln\like = &- \mu T \nonumber\\
&- \sum_{t_i} \frac{n}{Z} \left( \sum_{k=0}^M (a_k)^{1-p} (1-e^{-\frac{T-t_i}{a_k}}) - \frac{S}{a_{-1}} (1-e^{-\frac{T-t_i}{a_{-1}}} ) \right) \nonumber\\
&- \sum_{z_j} \frac{\alpha_N^{\C}}{\beta_N^{C}} (1-e^{-\beta_N^{\C}(T-z_j)} )\nonumber\\
&- \sum_{z_j} \frac{\alpha_N^{\NC}}{\beta_N^{\NC}} (1-e^{\beta_N^{\NC}(-z_j)})\nonumber\\
&+ \sum_{t_i} \ln \left[ \mu + \frac{n}{Z} \left( \sum_{k=0}^M (a_k)^{-p} R_k(i) - S\,R_S(i) \right) \right.\nonumber\\
&+ \left. \sum_{z_j<t_i} \alpha_N^{\C} e^{-\beta_N^{\C} (t_i-z_j)}+\sum_{z_j>t_i} \alpha_N^{\NC} e^{\beta_N^{\NC} (t_i-z_j)}\right].
\end{align}
The expression for the causal-only model is easily obtained from the above ignoring the terms which contain the non-causal terms. 

The results of this paper are obtained by using a quasi-Newton optimization routine for optimization. Since all the parameters are required to be positive or non-negative, we employed an optimizer that can accept box constraint (\textit{nlminb} of the package R). To increase the chances of finding the global optimum a set of 10 different starting points was tried, and the result with the highest likelihood was retained. The starting values were chosen so to span as much as possible the space of "reasonable" parameters values. In fact, most parameters have a physical interpretation that facilitates this choice. For instance, the baseline intensity $\mu$ cannot far exceed the average rate observed in the market.
We repeated the procedure for each pair of currencies and for both types of kernel. Uncertainties on the best parameters estimates are calculated using the inverse of the Hessian matrix at the maximum

\section{Comparison of simulated data from the models with and without the news term.} 
\label{sec:comparison_appendix}

\begin{figure}[]
\centering
\includegraphics[width=.45\textwidth]{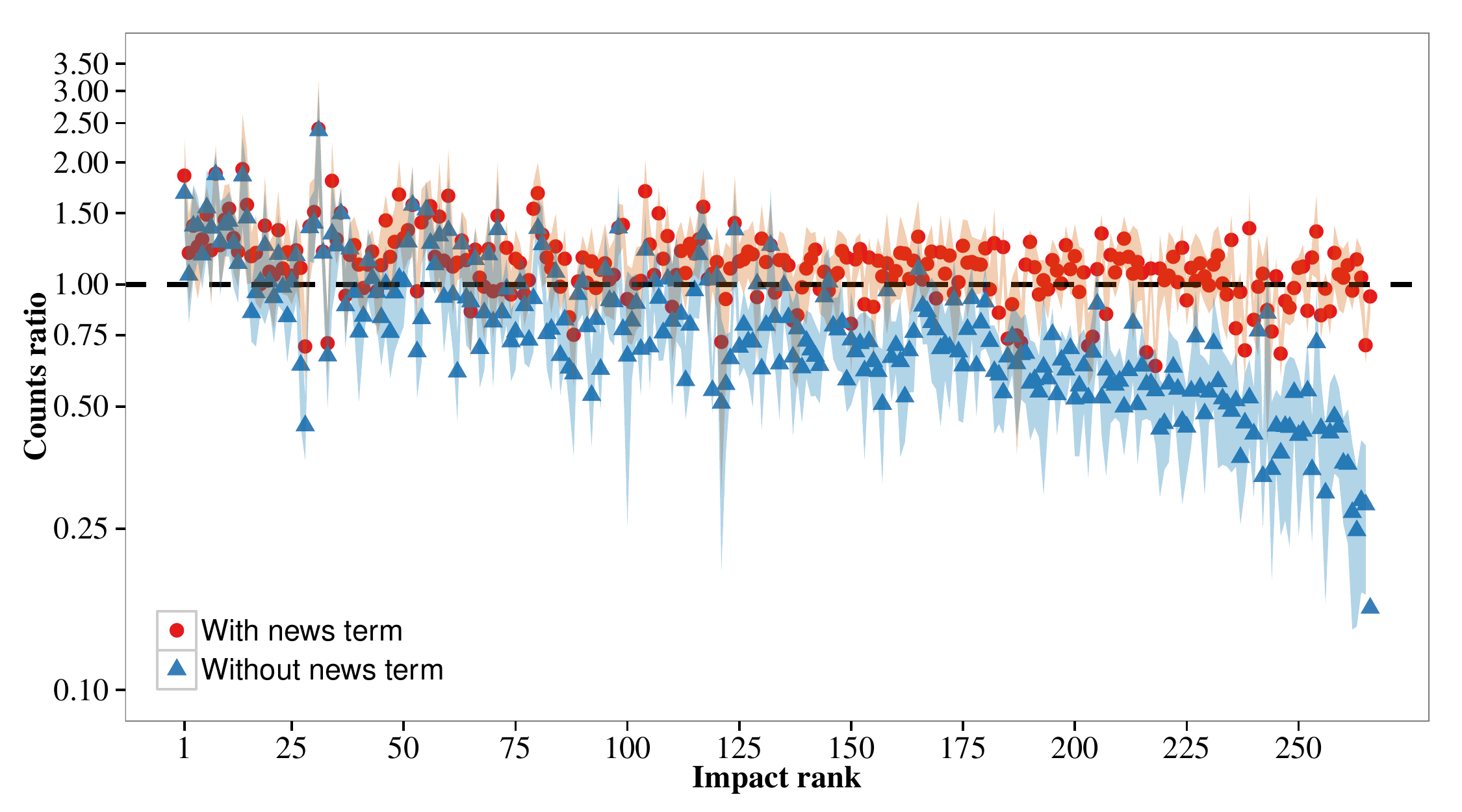}
\includegraphics[width=.45\textwidth]{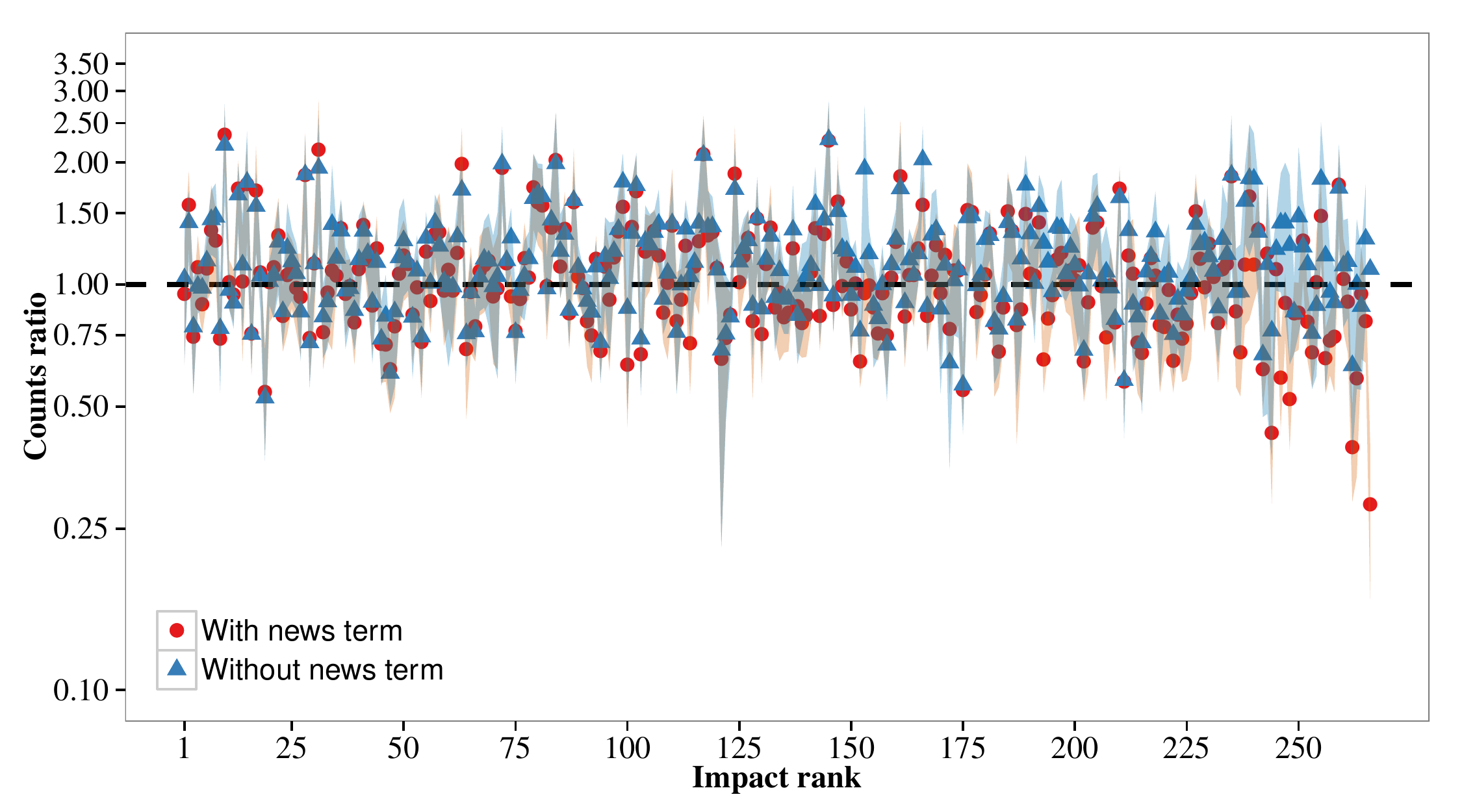}
\caption{(Color online). Ratio between mean of simulated counts and real counts in the 5 minutes window that follows (top panel) or precedes (bottom panel) the news plotted as a function of news' impact rank. Simulations were repeated 25 times for each news. All 266 news in our sample are considered. The figure shown refers to EUR/USD pair. A one standard deviation confidence interval is also plotted. }
\label{fig:ratio_vs_rank}
\end{figure}
In this appendix we compare further the Hawkes model with a power law endogenous kernel and an exogenous (news) term and the one without it.
For each of the 266 news examined and for each currency pair, we simulated the two models 25 times. We then count the number of events produced by each model's simulation in the 5 minutes that follow the news announcement. 
In the top panel of Figure \ref{fig:ratio_vs_rank} we plot the ratio between the mean of the simulated counts and the observed (real) counts 
$ \frac{<N_{\mbox{\tiny{sim}}}>}{N_{\mbox{\tiny{real}}}}$ in the 5 minute window that follows the news. The abscissa is the rank of the news in terms of its impact $\theta$ in ascending order. We observe that the two models are essentially equivalent for low impact news, whereas the model with the news term performs increasingly better as the impact grows. In fact, the model without the news term systematically underestimate the number of events after high impact news, since the self-exciting endogenous mechanism is unable to reproduce such a sudden increase in activity.
It is worth noting that the model with the news term does not artificially introduce a news effect when the news had little impact. For reference we also show in the bottom panel of Figure \ref{fig:ratio_vs_rank} the above ratio computed in the 5 minutes that precede the news. We observe that the two models produces almost identical outputs in this case.
\begin{figure}[h]
\centering
\includegraphics[width=.47\textwidth]{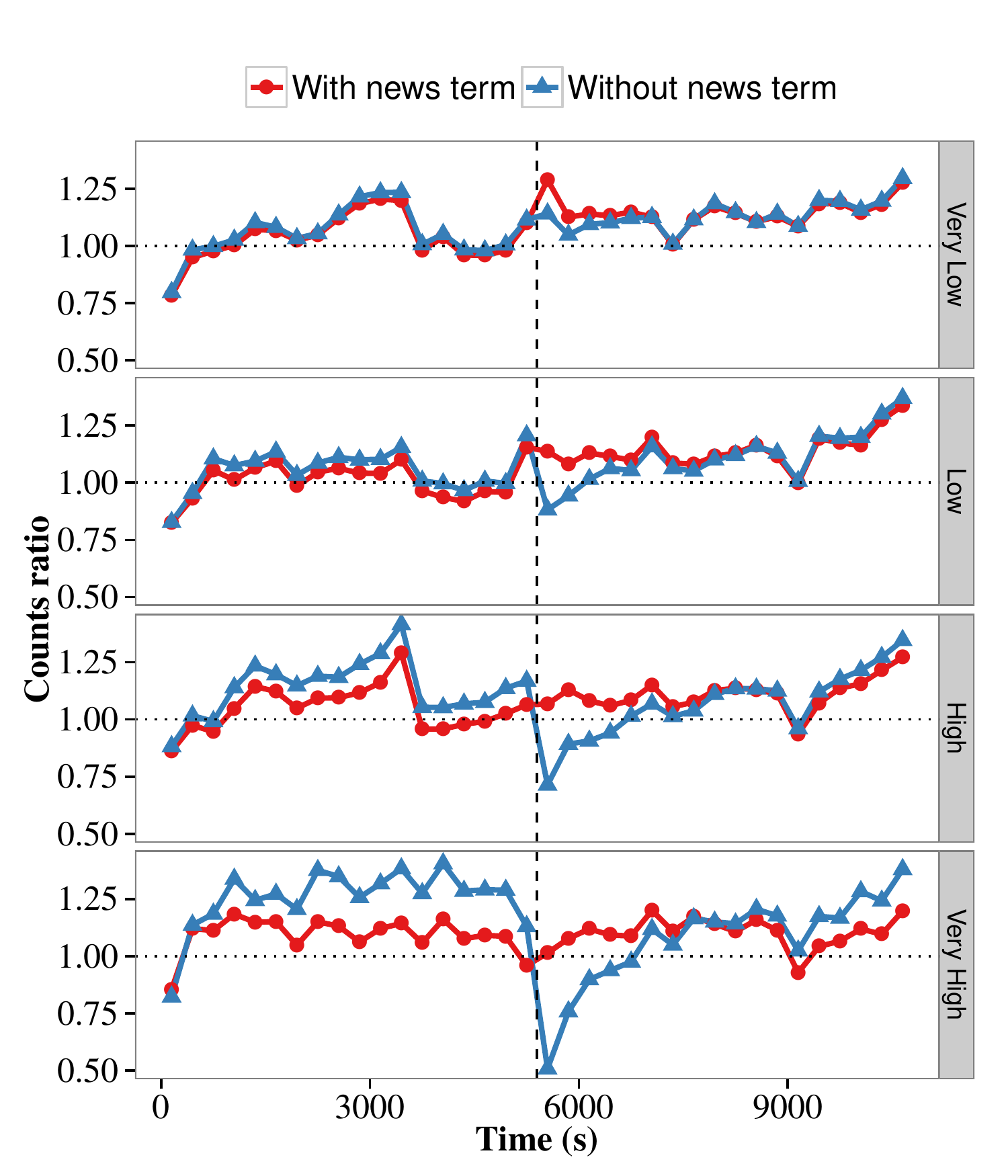}
\caption{(Color online). Ratio between mean of simulated counts and real counts computed in 5 minutes bin. Results form each bin are averaged across all the news in the same group of impact $\theta$. The plot refers to EUR/USD data.}
\label{fig:ratio_vs_time_quart}
\end{figure}

These considerations are reinforced by observing Figure \ref{fig:ratio_vs_time_quart}. The plot is built as follows: for each of the two models and for every news in the sample we divide the 3 hours window in bins of 5 minutes, and compute $ \frac{<N_{\mbox{\tiny{sim}}}>}{N_{\mbox{\tiny{real}}}}$ in each bin.
We also divide the news in four groups based on quartiles of the impact $\theta$, labeled as "Very Low", "Low", "High",and "Very High". We then take the average of the ratio across all news in the same group for each 5 minutes bin. In this way we measure the averave behavior predicted by the models as a function of time distance from the news announcement. Also in this plot we observe that the models behave in the same way for very low impact news. However, as soon as the news has a significant impact, the model without the news term strongly underestimates the activity that follows the news. Moreover, for very high impact news the model without the news term also systematically overestimate the activity before the news. This is because, without the exogenous kernel, the extra activity due to the news is attributed to the endogenous mechanism and thus it is spread over the full time interval.

\bibliography{biblio}
\end{document}